\documentclass[twocolumn,showpacs,aps,pra,amsmath,amssymb]{revtex4-1}
\usepackage{graphicx}
\usepackage{dcolumn}
\usepackage{bm}
\usepackage{hyperref}
\usepackage{subfigure}
\newcommand{\bra}[1]{\left\langle #1\right|}
\newcommand{\ket}[1]{\left|#1\right\rangle}
\begin{document}

\preprint{APS/123-QED}

\title{Quantum Coherence and Correlations of optical radiation by atomic ensembles interacting with a two-level atom in microwave cavity}

%
\author{\"{O}. E. M\"{u}stecapl{\i}o\u{g}lu}
\email{omustecap@ku.edu.tr}
\affiliation{ Department of Physics, Ko\c{c} University, Sar{\i}yer,
Istanbul, 34450, Turkey} 
\affiliation{ Institute of Quantum
Electronics, ETH Zurich, 8093 Zurich, Switzerland}
%
\date{\today}

\begin{abstract}
We examine quantum statistics of optical photons emitted from 
atomic ensembles which are classically driven and simultaneously coupled to 
a two-level atom via microwave photon exchange. Quantum statistics and correlations are analyzed by calculating second order coherence degree, von Neumann entropy, spin squeezing for multi-particle entanglement, as well as genuine two and three-mode entanglement parameters for steady state and non-equilibrium situations. Coherent transfer of population between the radiation modes and quantum state mapping between the two-level atom and the optical modes are discussed. 
A potential experimental realization of the theoretical results in a superconducting coplanar waveguide resonator containing diamond crystals with Nitrogen-Vacancy color centers and a superconducting artificial two-level atom is discussed.
\end{abstract}


\maketitle

\section{Introduction}\label{sec:intro}

Primary requirements of successful quantum information technology, in particular of large scale distributed quantum networking, is to have efficient and reliable means of storage \cite{qMemory}, processing \cite{qProcessor} and dissemination \cite{qComm} of quantum resources. These tasks can most likely be solved in different modules, designed and optimized for their ideal operation. The composite quantum information device would be then a hybrid structure of such sub-systems \cite{hybridDeviceReview}. As the individual modules can have different time and energy scales of operation, the fundamental challenge is to merge them together congruously.  

In this paper we consider an atomic three-level ensemble, with the so called $\Delta$-type level scheme \cite{DeltaScheme,cyclic_ensemble_model}. We argue that such a system can be a promising set up for controllable exchange of quantum resources between energetically remote modules in a hybrid quantum information device. Basic idea is to exploit the presence of an extra energetic component in the $\Delta$-scheme to bridge the large energy gap between the other two components. In particular we envision a typical case where one component is a two-level atom, corresponding to the stationary quantum hardware and quantum information unit (qubit). It is coupled to the $\Delta$ system by a cavity field in the microwave domain. The other component could be a remote quantum communication node such as a satellite or quantum repeater along an optical fiber so that these require coupling to the $\Delta$-system in the optical domain. Hence by choosing the extra available transition also in the optical domain, we can consolidate these two domains. The $\Delta$ system could play a double faceted role. It can serve as the quantum memory in terms of the pure or, together with the two level atom, hybrid spin ensemble qubits \cite{hybridSpinEnsembleQubit}. In addition, it becomes a mediator of communication between the stationary quantum hardware and the quantum communication channels for optical flying qubits interconnecting remote nodes of a quantum network.  

Due to dipole selection rules, $\Delta$-type transitions can happen only in certain systems, either lacking, or with explicitly broken, inversion symmetry, such as artificial atoms \cite{DeltaScheme,DeltaArtifAtom}, semiconductors \cite{DeltaSemiCond1,DeltaSemiCond2,DeltaSemiCond3} or chiral molecules \cite{DeltaChiral1,DeltaChiral2}. Alternatively, one of the transitions can be of magnetic dipole or electric quadrupole type in the $\Delta$-scheme \cite{magneticdipoleDelta1,magneticdipoleDelta2,magneticdipoleDelta3}. Such weak transitions are collectively enhanced for strong coupling due to the ensemble \cite{hybridSpinEnsembleQubit}. We specifically consider the latter case of magnetic dipole assisted $\Delta$-scheme, though our results would be valid in other cases, too. 

Some quantum statistical properties of single isolated $\Delta$-type atomic ensembles have been explored recently \cite{cyclic_ensemble_model,DeltaSchemeQCorrelations}. Coupling of atomic ensembles, in particular, three-level $\Lambda$-type ensembles to a two-level atom has been discussed in the context of quantum state mapping from quantum memories to charge qubit \cite{chargeQubit} in transmission line resonators \cite{cpwg_exp1,cpwg_exp2,cpwg_exp3,qMemTLR}. In this paper we examine quantum coherence and correlations of the optical photons emitted by a $\Delta$-type ensemble coupled to a two-level atom in a microwave cavity. We also investigate the case of two $\Delta$-type ensembles. Equilibrium and non-equilibrium situations are separately examined. Conditions of photon antibunching, particle entanglement \cite{spinsqzParam1,spinsqzParam2}, genuine bi-partite and tri-partite mode entanglement, W state \cite{Wstate} are revealed. Furthermore, quantum state mapping between the two-level atom and optical modes, as well as coherent transfer of population between the radiation modes are found.

This paper is organized as follows. In Sec. \ref{sec:model_sys} we describe the model system and by eliminating the collective atomic states, we derive the effective Hamiltonians for two-level atom and radiation modes. In Sec. \ref{sec:results_discussions}, we present our results and discussions for the single and two-ensemble cases in the steady state and non-equilibrium situations. In Sec. \ref{sec:expProposal}, a potential experimental system for physical implementation of our results is described. Specifically, we consider an extension of recently realized Nitrogen-Vacancy (N-V) centers in diamond crystals strongly coupled to superconducting coplanar waveguide resonators (CPWG) \cite{NV_CPWG_exp}. Similar electron spin ensemble to on-chip superconducting cavity couplings, specifically coupling of  Cr$^{3+}$ spins in ruby and N substitution (P1) centers in diamond, are also observed in parallel efforts \cite{rubyDiamondExp}. Finally in Sec. \ref{sec:concl}, we conclude. 
\section{$\Delta$-type Atomic Ensemble coupled to a two-level atom}
\label{sec:model_sys}
\subsection{Model Hamiltonian}

\begin{figure}[!tbp]
\begin{center}
\includegraphics[width=8.6 cm]{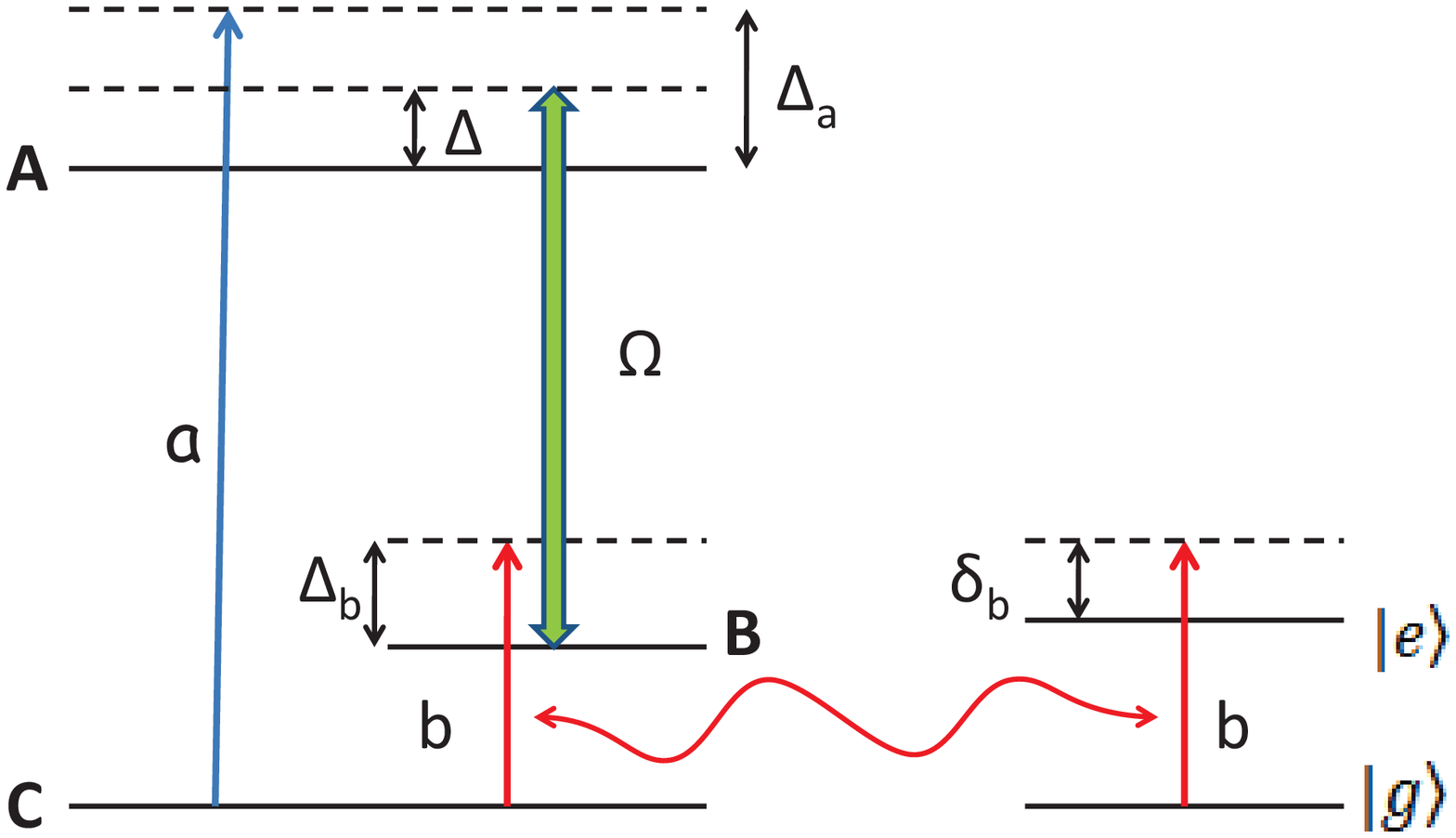}
\caption{\label{fig:level_scheme}(Color online) Identical three-level atoms in the ensemble interact with two quantized (indicated by $a,b$) and one classical (indicated by $\Omega$) 
electromagnetic fields in $\Delta$ scheme. The microwave photons further couple the atomic ensemble to a single two-level atom. The classical drives on C-A transition and on the cavity field are not shown.}
\end{center}
\end{figure}

We consider a collection of $N$ identical three-level atoms interacting with three electromagnetic fields in $\Delta$-type transition scheme delineated in Fig. \ref{fig:level_scheme}. The atomic ensemble is further coupled to a two-level atom through microwave photon exchange. The lowest level $\mid C\rangle$ is coupled to the intermediate 
level $\mid B\rangle$ by a magnetic dipole transition of rate $g_b$, 
while the transitions
from the highest level $\mid A\rangle$ to the lower doublet  $\mid B\rangle$
and  $\mid C\rangle$ are optical electric-dipole transitions of rates 
 $\Omega$ and $g_a$, respectively. We further suppose that optical fields are traveling waves while the microwave field is a standing wave of a cavity that contains the two-level
atom and the atomic ensemble.

Our model can be conceived as a modest generalization of the case of a $\Delta$-type atomic ensemble, resonantly coupled to the optical fields, which is studied for the bi-partite entanglement of the quantized fields \cite{cyclic_ensemble_model}. On the other hand, such a seemingly innocent addition of two level atom generalizes
the system from two interacting linear oscillators to the case of a linear oscillator coupled to a nonlinear oscillator, and as such can cause remarkable changes in the quantum statistical properties of the fields. 
In comparison to $\Lambda$-type ensemble coupled to two-level atom \cite{darkstate}, $\Delta$-type scheme allows for effective interaction between $a,b$ modes. In addition, we take into account additional external drives and consider the system as open with all the decay channels due to cavity losses or spontaneous emission decays, which causes subtle effects in particular in the steady state.    

The system can be described by a model Hamiltonian $H=H_0+H_{1}+H_{p}$
where (in units of $\hbar$)
\begin{eqnarray}\label{eq:mainModel}
H_0&=& \sum_{\vec{k}}\omega_{a\vec{k}} a_{\vec{k}}^\dag a_{\vec{k}}+
\omega_b b^\dag b+\frac{\omega_0}{2}\sigma_z+\sum_{xj}\omega_xR_{xx}^{(j)}, \\
H_{1}&=&g\sigma b^\dag+
\Omega\sum_{j}R_{AB}^{(j)}\mathrm{e}^{i(\vec{k_d}\cdot\vec{r}_j-\omega_dt)}
\nonumber\\
&&+\sum_{j,\vec{k}}g_{a\vec{k}}\mathrm{e}^{i\vec{k}\cdot\vec{r}_j}R_{AC}^{(j)}a
+g_b\sum_{j}R_{BC}^{(j)}b 
+h.c.. \\
H_p&=&\Omega_A\sum_jR_{AC}^{(j)}
\mathrm{e}^{i(\vec{k_A}\cdot\vec{r}_j-\tilde\omega_At)}
+E_b\mathrm{e}^{-i\tilde\omega_bt}b+h.c.
\end{eqnarray}
Here, $x=A,B$,$j=1,..,N$, $\omega_{a\vec{k}},\omega_b,\omega_{A},\omega_B,\omega_0$ are 
respectively the frequencies of the electromagnetic fields and atomic transitions relative to the lowest level $\mid C\rangle$. Probe field frequencies are denoted by $\tilde\omega_A$ and $\tilde\omega_b$. Effective drive strength of the microwave cavity and the Rabi frequency of the drive for the $\ket{C}$ to $\ket{A}$ transition are represented by $E_b$ and $\Omega_A$, respectively.
Population and transition operators of single atoms in the ensemble are respectively represented by $R_{xx}^{(j)}$ and $R_{xy}^{(j)},x\neq y$ with
$x,y=A,B,C$. Optical and microwave photon annihilation (creation) operators denoted by $a_{\vec{k}},b$ ($a_{\vec{k}}^\dag,b^\dag$), respectively. 
The coupling coefficients
are assumed to be spatially homogeneous. We suppose that the 
wave vectors of the optical fields satisfy the mode matching
condition \cite{cyclic_ensemble_model} $\vec{k}=\vec{k}_d=\vec{k}_A$ 
so that quantized optical field can be treated
as single mode. More detailed conditions on three-dimensional analysis of mode matching between free space light and the atomic ensemble will be ignored for simplicity \cite{modematching}. Suppressing the constant global phase factor by  
\cite{phase_transform} $\mid A\rangle\rightarrow \exp{(i\vec{k_d}\cdot\vec{r}_j)}\mid A\rangle$, we drop the $\vec{k}$ index from our notations and further considerations.  

There would be additional Langevin noise terms in the Hamiltonian, corresponding to the coupling to the continuum of other modes. These will be taken into account effectively in the master equation formalism and shall not be shown here.
\subsection{Quasi-spin wave picture of the model system}

It is illuminating to portray the model system in terms of collective excitation
operators \cite{collectiveExcOps} 
$A=X_{CA}/\sqrt{N},B=X_{CB}/\sqrt{N},
T=X_{BA}$,
where 
$X_{xy}=\sum_{j=1}^NR_{xy}$,
is analogous to the Hubbard operators \cite{hubbard} in strongly correlated
systems. In the ensemble, the set of Hubbard operators represent SU(3) symmetry of a spin-1 system. They can be associated with the Gell-Mann fundamental representation of the SU(3) Lie algebra by choosing $U=X_{CA},V=X_{BC}$ and $T=X_{BA}$, so that $U,V,T$ form the three spin-$1/2$ subgroups. Each obeys the SU(2) algebra with $[L,L^\dag]=-2L_3,L=U,V,T$. They are interdependent by the commutations such as $[V^\dag,U^\dag]=-T^\dag$. Analogous to excitons and magnons, SU(2) algebras of U and V sub-spins can be contracted to Weyl-Heisenberg algebra $h_2$ to associate collective excitations with quasi-spin wave quanta \cite{collectiveExcOps}. 
The system space becomes a semidirect product Lie algebra, $SU(2)\bar\otimes h_2$. 
In the limit of low excitations to
$\mid A\rangle$ and $\mid B\rangle$, relative to $N$,  $V_3\approx N/2$ and $U_3\approx -N/2$. With the contraction parameter $1/\sqrt{N}$, the operators $A=U/\sqrt{N}$ and $B=V^{\dag}/\sqrt{N}$ become  
approximately bosonic. They become independent, as the isospin subgroup is weakly populated. The isospin is consistently represented by $T=B^\dag A$.

Denoting the phases of the $g,g_a,g_b,\Omega,\Omega_A,E_b$ by $\theta_g,\theta_a,\theta_b,\theta_\Omega,\theta_A,\tilde\theta_b$, respectively, they can be suppressed by $b\rightarrow b\exp{(i\tilde\theta_b)},\sigma\rightarrow \sigma\exp{[i(\theta_g-\tilde\theta_b)]}, B\rightarrow B\exp{[-i(\theta_b+\tilde\theta_b)]},a\rightarrow a\exp{[i(\theta_a+\theta_A)]}$ and $A\rightarrow A\exp{(-i\theta_A)}$. Employing a unitary 
transformation by $U=exp{(-i\omega_dt)}$ with $H\rightarrow U^\dag HU-iU^\dag \partial_tU$, the Hamiltonian becomes
\begin{eqnarray}
\label{eq:bosonModelH0_rf}
H_0&=&\sum_{x=a,A}\varpi_xx^\dag x+\omega_bb^\dag b+\omega_BB^\dag B+
\frac{\omega_0}{2}\sigma_z,\\
\label{eq:bosonModelH1_rf}
H_1&=&g\sigma b^\dag+g_{a} A^\dag a+ g_{b} B^\dag b+
\Omega \mathrm{e}^{i\phi}A^\dag B +h.c.,\\
H_p&=&\Omega_A\mathrm{e}^{-i\tilde\varpi_At}A^\dag+
E_b\mathrm{e}^{-i\tilde\omega_bt}b+h.c.
\end{eqnarray}
where the coupling coefficients $g_{a,b} \rightarrow \sqrt{N}g_{a,b}$ are collectively enhanced; $\phi=\theta_\Omega+\theta_A-\theta_b-\tilde\theta_b$, $\varpi_{a,A}=\omega_{a,A}-\omega_d$ and $\tilde\varpi_{A}=\tilde\omega_{A}-\omega_d$. Even though it can play curious role on quantum correlations \cite{cyclic_ensemble_model}, for simplicity we take $\phi=0$. We further assume three-photon resonance condition 
$\tilde\varpi_A=\tilde\omega_A-\omega_d=\tilde\omega_b$ for the classical driving fields. 
\subsection{Effective Hamiltonian for the model system}
$\Delta$-type ensemble can induce an effective photon - photon interaction mediated by the quasi-spin wave background \cite{cyclic_ensemble_model}. In order to show that explicitly, let us introduce the normal modes $p_{\pm}=u_\pm A+v_\pm B$ by the Bogoliubov transformation \cite{bogoliubov} 
with the relations $u_-=-v_+$ and $v_-=u_+$, where $u_\pm=\sqrt{(\eta\pm(\varpi_A-\omega_B))/2\eta}$ and $\eta=\sqrt{(\varpi_A-\omega_B)^2+4\Omega^2}$.

The model Hamiltonian can be rewritten as $H=H_0+H_1+H_p$ where
\begin{eqnarray}
\label{eq:polaritonAB_H0}
H_0&=&\varpi_aa^\dag a+\omega_bb^\dag b+\sum_{\lambda=\pm}\Omega_\lambda p_\lambda^\dag p_\lambda +
\frac{\omega_0}{2}\sigma_z,\\
\label{eq:polaritonAB_H1}
H_1&=&g\sigma b^\dag+\sum_{\lambda=\pm}p_{\lambda}^\dag
(g_{a\lambda} a + g_{b\lambda} b)+h.c.,\\
H_p&=&\sum_{\lambda=\pm}
\Omega_{A\lambda}\mathrm{e}^{-i\tilde\varpi_At}p_\lambda
+E_b\mathrm{e}^{-i\tilde\omega_bt}b+h.c.
\end{eqnarray}
Here $\Omega_\pm=(\varpi_A+\omega_B\pm\eta)/2$, $g_{a\lambda}=g_au_{\lambda},g_{b\lambda}=g_bv_{\lambda}$ and 
$\Omega_{A\lambda}=\Omega_Au_{\lambda}$.

Effective coupling between the electromagnetic modes can be obtained by
a unitary transformation to eliminate the interactions between the electromagnetic and quasi-spin-wave modes. We introduce an anti-Hermitian operator $S$ 
\begin{eqnarray}
S=\sum_{\lambda=\pm}(x_{\lambda}a+y_{\lambda}b)p_\lambda^\dag-h.c.,
\end{eqnarray}  
where the unknown coefficients $x_{\lambda}$ and $y_\lambda$ are to be determined by the Fr\"ochlich - Nakajima canonical transformation condition \cite{frochlich1,frochlich2,frochlich3}
\begin{eqnarray}
[H_0,S]=-\sum_{\lambda=\pm}p_{\lambda}^\dag
(g_{a\lambda} a + g_{b\lambda} b)+h.c.,
\end{eqnarray}
which yields $x_\lambda=g_{a\lambda}/(\varpi_a-\Omega_\lambda)$ and
$y_\lambda=g_{b\lambda}/(\omega_b-\Omega_\lambda)$.

Assuming $\Omega\gg g_a,g_b$, an effective Hamiltonian $H_{eff}=\exp{(-S)}H\exp{(S)}$ can be written up to the second order in $x_\lambda,y_\lambda$,
\begin{eqnarray}
\label{eq:effectiveHfull}
H_{eff}&=&\varpi_a^\prime a^\dag a+\omega_b^\prime b^\dag b+\frac{\omega_0}{2}\sigma_z+g(\sigma b^\dag+h.c.)\nonumber\\
&+&(E_b^\prime\mathrm{e}^{-i\tilde\omega_bt}b
+E_a\mathrm{e}^{-i\tilde\varpi_At}a^\dag-Ja^\dag b + h.c)\nonumber\\
&+&\sum_{\lambda}\Omega_{\lambda}^\prime p_\lambda^\dag p_\lambda+\sum_{\lambda\neq\mu}Q_{\lambda\mu} (p_\lambda^\dag p_\mu+h.c.)\nonumber\\
&+&\sum_{\lambda}(G_{\lambda}\sigma p_\lambda^\dag+
E_{p\lambda}\mathrm{e}^{-i\tilde\omega_bt}p_\lambda^\dag+h.c.)
\end{eqnarray}
with $\lambda,\mu=\pm$. The coefficients in $H_{eff}$ are listed in the appendix \ref{app:effectiveH_coeffs}. The indirect coupling between the quasi-spin-wave modes and the electromagnetic modes is through their common interaction with the two-level atom. In the large $\Omega$ limit, the quasi-spin waves are strongly off-resonant with the two-level atom and practically uncoupled. More formally, in the dispersive regime,  $\Omega-\omega_0\sim\Omega\gg g$, employing another Fr\"ochlich transformation, such an interaction can be reduced to a Stark shift of two-level atom resonance proportional to the population of the quasi-spin normal modes, that can be neglected for weak excitations. Influence of such sequential canonical transformation on radiation modes would be negligible provided that $\Omega\gg g_a\gg g_b,g$. In addition, effective weak drive on the two-level atom would be negligible. The effective Hamiltonian, keeping the leading terms only, is  reduced to
\begin{eqnarray} \label{eq:single_ensemble_heff}
H_{eff}&=&(\varpi_a-\tilde\omega_b) a^\dag a+(\omega_b-\tilde\omega_b) 
b^\dag b+\frac{\omega_0-\tilde\omega_b}{2}\sigma_z \nonumber\\
&&+g(\sigma b^\dag+h.c.)-J(a^\dag b+h.c.)\nonumber\\
&&+E_a(a^\dag+a)+E_b(b^\dag+b).
\end{eqnarray}
Here, three-photon resonance condition for the driving classical fields is employed and the Hamiltonian is written in the frame rotating at $\tilde\omega_b$. We further suppose that three-photon resonance holds for the radiation fields 
such that $\omega_a=\omega_d+\omega_b$, then $\varpi_a-\tilde\omega_b
=\omega_b-\tilde\omega_b\equiv\delta$ is defined. We find $\omega_0-\tilde\omega_b=\delta-\delta_b$. According to Fig. \ref{fig:level_scheme}, $\delta_b=\omega_b-\omega_0,\Delta=\omega_d-\omega_{AB},
\Delta_a=\omega_a-\omega_A,\Delta_b=\omega_b-\omega_B$, in terms of which we express the three-photon resonance condition $\Delta_a-\Delta-\Delta_b=0$.

The Hamiltonian of Eq. \ref{eq:single_ensemble_heff} has a zero-eigenvalue solution in the single excitation manifold, or the so-called dark state,  which is given by \cite{darkstate}
\begin{eqnarray}
\mid DS\rangle=(\cos{\alpha}\sigma-\sin{\alpha}a)^\dag\mid vac\rangle,
\end{eqnarray}
with $\alpha=-\arctan{(g/J)}$
Presence of such a dynamically fixed state has been exploited to propose an adiabatic transfer
protocol to map the quantum information stored in a molecular $\Lambda$ type
ensemble to a charge qubit of two-level atom, and vice versa. Our treatment
of cyclic ensemble reveals a complement to this protocol by allowing quantum
information transfer from two-level atom stationary qubit to optical flying qubit. A state $(x\ket{g}+y\ket{e})\ket{00}_{ab}$ can be adiabatically transferred into $(x\ket{0}_a+y\ket{1}_a)\ket{g}\ket{0}_b$. The decoherence channels due to two-level atom and the microwave cavity are suppressed. The effective loss channel of the optical photon, transferred from the spontaneous emission of the level $\ket{A}$ during the Fr\"ochlich transform, is compensated with the effective drive $E_a$. Taking into account time dependent interaction coefficients and time-dependent canonical transformations, an optimal fidelity of the state transfer can be characterized. 

Alternative to the derivation based upon sequential Fr\"ochlich transformations as presented here, it is possible to get the effective Hamiltonian with a single step Fr\"ochlich transformation or by reducing a general four-color $(a,b,A,B)$ polariton to a two color $(a,b)$ one \cite{cyclic_ensemble_model} for strong drive $\Omega$ conditions. It is straightforward to generalize this model by either of these methods to the case of a pair of cyclic atomic ensembles in the microwave cavity, coupled to the same two-level atom. We find
\begin{eqnarray} \label{eq:double_ensemble_heff}
H_{eff}&=&\sum_{j=1,2}\varpi_{aj}^\prime a_j^\dag a_j+\omega_b^\prime b^\dag b+\frac{\omega_0}{2}\sigma_z \nonumber\\
&&+g(\sigma b^\dag+h.c.)-\sum_{j=1,2}J_j(a_j^\dag b+h.c.)\nonumber\\
&&+\sum_{j=1,2}E_j(a_j^\dag+a_j)+E_b(b^\dag+b).
\end{eqnarray}
For identical symmetric ensembles,  $\varpi_{aj}=\varpi_a$,$E_j=E_a$ and $J_j=J$.

The single ensemble effective model reduces to the Jaynes-Cummings (JC) model for the optical mode in the dispersive limit of the microwave cavity.  Two-ensemble model becomes
a two-site Bose-Hubbard (BH) model,
with pure and cross-Kerr interactions of the type $(a_i^\dag a_i)^2,i=1,2$ and $a_1^\dag a_1a_2^\dag a_2$. A common proposal to realize a BH dimer is to use JC-BH systems of tunnel coupled nonlinear cavities, where the nonlinearity
is provided by a two-level atom in each of them \cite{JCBH,JCDimer}. In contrast to such 
direct methods to realize atom-optical analogs of Josephson networks, 
here there is only one two-level atom providing shared nonlinearity 
by each optical mode, which simplifies architecture of identical homogeneous coupled subsystems. 

BH dimer can be described as a pseudo-spin system with quadratic nonlinearities using pseudo-spin Schwinger operators $S_+=a_1^\dag a_2, S_z=(a_1^\dag a_1-a_2^\dag a_2)/2$. When sufficiently strong nonlinearity present, it leads to multi-particle entanglement between the distant optical modes via the so-called axis twisting spin squeezing route \cite{spinsqzParam1,spinsqzParam2}. The question of whether multi-particle entanglement can still be found in the non-dispersive regime is not as obvious to answer due to lack of immediate mapping onto a spin model. We shall explore that case numerically in the following section, together with other types of quantum correlations as well as with quantum statistical properties of radiation modes in the effective models.

Generalization of these models beyond two-ensembles is in principle possible. However, the quasi-spin wave mediated effective tunnel coupling here is long range beyond nearest neighbors and hence direct relation to simple Bose-Hubbard models is not immediate. 


\section{Results and Discussions}\label{sec:results_discussions}

Assuming the effective model remains Markovian and Born-Markov approximation is applicable, dynamics can be investigated by a quantum master equation  
\begin{eqnarray}
\dot\rho=-i[H,\rho]+\sum_{x}\kappa_x{\cal D}[x]\rho+\gamma{\cal D}[\sigma]\rho
+\frac{\gamma_\phi}{2}{\cal D}[\sigma_z]\rho.
\end{eqnarray}
Here $\kappa_x$ is the photon loss rate of the mode $x=a,b$, $\gamma_1$ and $\gamma_\phi$ are respectively the relaxation and pure dephasing rates of the two-level atom. Dissipation terms ${\cal D}[x]\rho=(2x\rho x^\dag-x^\dag x\rho
-\rho x^\dag x)/2$ are the Liouvillian superoperators for damping in Lindblad form. 

Fr\"ochlich transformations used in the derivations of the effective models influence the decay rates. In the leading order, we estimate $\kappa_a\sim\gamma_A(g_a/\Omega)^2$, where $\gamma_A$ is the excited state decay rate of the $\Delta$-type atom. We choose $g_a/\Omega=0.1$ so that $\kappa_a=0.01\gamma_A$. We assume $\kappa_a=\kappa_b$. In addition, we suppose
$E_a=E_b=0.1\kappa_b$, which corresponds to weak driving condition for which empty cavity photon number is $0.01$ on resonance. Such strict equalities are to reduce the number of independent parameters in the simulations but not crucial in practice. It is sufficient to arrange field-atom detunings and the driving field strengths (The formulae in the Appendix \ref{app:effectiveH_coeffs} can be referred) to ensure sufficient effective drive $E_a$ can be translated from $A$ to $a$ so that the optical radiation can survive in the steady state against the decay channels.

We numerically solve the master equation using the "Quantum Optics Toolbox" \cite{qotoolbox} and determine the steady state and dynamical behavior of the system. Our typical results will be reported below for the cases of single and two-atomic ensembles separately. For all cases we fix  $J/2\pi=1$ MHz. 

\subsection{Single Atomic Ensemble}
\subsubsection{Steady State Coherence and Correlations} 

Taking $\delta_b=J$, $\kappa_b=\kappa_a=(2\pi)0.4$ MHz,
$\gamma_1/2\pi=0.02$ MHz, $\gamma_\phi/2\pi=0.3$ MHz, steady state solution of the master equation is determined, in a truncated Fock space of $3$ photons in each mode, for a range of $g/2\pi\in \{0,4\}$ MHz and 
$\delta / 2\pi\in \{-5,5\}$ MHz. 

\begin{figure}[!tbp]
\begin{center}
\includegraphics[width=8.6 cm]{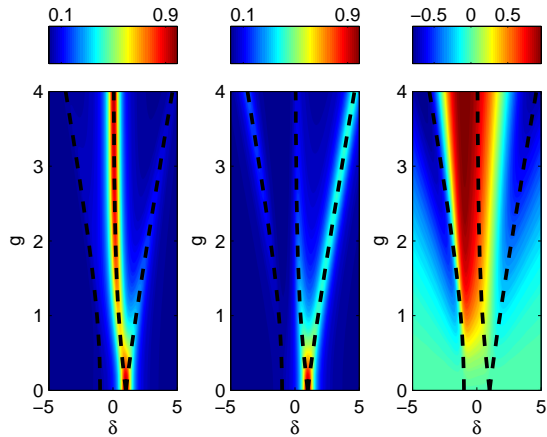}
\caption{\label{fig:transmissions}(Color online) Steady state populations $n_a$ (Left Figure), $n_b$ (Middle Figure), and their relative fractional population $n_{ab}=(n_a-n_b)/(n_a+n_b)$ (Right Figure) of the optical and microwave modes at $J/2\pi=1$ MHz,  $\delta_b=J$, $\kappa_b=\kappa_a/=(2\pi)0.4$ MHz, $\gamma_1/2\pi=0.02$ MHz, $\gamma_\phi/2\pi=0.3$ MHz. Fock spaces for the radiation modes are taken to be four dimensional. Dashed lines indicate the probe resonances with the single excitation manifold. $g/2\pi$ and $\delta/2\pi$ are in MHz units.}
\end{center}
\end{figure}

Populations of the radiation modes are shown in Fig. \ref{fig:transmissions} in the  $(\delta,g)$ phase space. Dashed lines indicate the probe resonances with the single excitation manifold. The relatively low transmission along a particular resonance is due to our choice of equal drive strengths. For a simple illustration, let us consider $\delta_b=0$ for which
\begin{eqnarray}
q_D&=&\frac{-g}{K}a-\frac{J}{K}\sigma,\\
q_\pm&=&\frac{1}{\sqrt{2}K}(-Ja\pm Kb+g\sigma),
\end{eqnarray}
are the dressed atom-polariton operators \cite{atomPolariton} of the single excitation manifold. For our choice of $\delta_b=0$, $q_D^\dag\ket{vac}$ is the dark state of the manifold. The states generated by the $q_\pm^\dag$ are with $\pm K$ energies where $K=\sqrt{g^2+J^2}$. For $g<J$ we find $a=(q_++q_-)/\sqrt{2}$ and $b=(q_++q_-)/\sqrt{2}$ so that the combined drive terms for the $a,b$ modes becomes $\sqrt{2}E_a(q_++q_+^\dag)$ if $E_a=E_b$. In that case neither $q_-$ nor $q_D$ are driven and hence they cannot survive to steady state. Fig. \ref{fig:transmissions} confirms with this argument as it exhibits that only bright polariton branch is the upper one, while the lower two remains in the dark. With the increase of $g$, $a$ mode drives mainly the $q_D$ polariton. This is seen in the Fig. \ref{fig:transmissions} as the bright middle polariton branch. The upper branch of $q_+$ is more strongly driven relative to $q_-$ and correspondingly associated with the bright $b$ line in the middle figure of Fig. \ref{fig:transmissions}. One can selectively populate different polariton branches, for example taking $E_b=-E_a$ yields bright $b$ mode along the lower polariton branch. 

The right figure in Fig. \ref{fig:transmissions} is the behavior of population fraction defined to be $w=(n_a-n_b)/(n_a+n_b)$, with 
$n_x=\langle x^\dag x\rangle, x=a,b$. As long as $g<J$ there is no distinct radiation phases in the $(\delta,g)$ space. When $g>J$ however, radiation in the system is dominated by the microwave mode along the upper and lower polariton branches. In between, there is a wedge like region about the middle polariton branch, where the radiation is dominated by the optical mode. The main effect of non-zero $\delta_b$ would be to change the slopes of the probe resonance lines,and as such the widths of the optical and microwave phases, but the above descriptions would be valid. 

We characterize quantum statistical coherence properties of the radiation modes in the steady state in terms of the second order coherence function at zero time delay,
\begin{eqnarray}
g^2(0)=\frac{\langle a^\dag a^\dag a a\rangle}{\langle a^\dag a\rangle^2}, 
\end{eqnarray}
the von Neumann entanglement entropy \cite{qProcessor},
\begin{eqnarray}
S=-Tr(\rho_a\log_2\rho_a),
\end{eqnarray}
and the genuine two-mode entanglement parameter \cite{GenuineModeEnt},
\begin{eqnarray}
\lambda_{ab}=\mid \langle a^\dag b\rangle\mid^2-\langle a^\dag ab^\dag b\rangle.
\end{eqnarray}  
Here, $\rho_a=Tr_b(Tr_\sigma(\rho_{ss}))$ is the reduced density matrix of optical mode $a$ evaluated by tracing out atomic ($\sigma$) and microwave ($b$) degrees of freedom. $S$ is an entanglement measure for pure bipartite states. As we have three sub-systems, we can interpret it as a measure of impurity in the steady state, analogous to linear entropy. On the other hand, we find that the two-level atom is weakly excited everywhere in the $\delta,g$ phase space for the steady state. Thus, $S$ bears the signs of profound quantum correlations between the radiation modes, especially in the $g<J$ regime where it should be a reliable measure of mode entanglement. A more clear signature of genuine two-mode entanglement can be found when $\lambda_{ab}>0$.  

Mode entanglement is due to correlations in the occupation number space in the second quantization framework. In the first quantization or in the particle picture, scattering of particles can lead to multi-particle entanglement. Though we do not have such direct particle interactions for photons, effective interactions that lead to photon blockade effect \cite{photonBlockade,photonblockade2} can yield similar multi-particle entanglement of photons. We want to explore if increased particle correlations among the photons can be found with the increasing anharmonicity parameter $g$ in regimes of sub-Poisson statistics. To characterize particle entanglement, we use spin squeezing parameter \cite{spinsqzParam1,spinsqzParam2}, generalized to the case of dissipation, \cite{spinsqzDissipation} defined to be
\begin{eqnarray}
\xi_x^2=\frac{\langle N\rangle 
[\Delta J_x]^2}{\langle J_y\rangle^2+\langle J_z\rangle^2}.
\end{eqnarray}
When $\xi_x^2<1$ the state is said to be particle entangled, so that it has non-separable density matrix in the first quantization. Here, the pseudo-spin operators $J_{x,y,z}$ are the usual Schwinger representation for the two bosonic modes $a$ and $b$,
\begin{eqnarray}
J_x&=&\frac{1}{2}(a^\dag b+ab^\dag),\\
J_y&=&\frac{1}{2}(a^\dag b-ab^\dag),\\
J_z&=&\frac{1}{2}(a^\dag a-b^\dag b).
\end{eqnarray}
$\langle N\rangle=\langle a^\dag a+ b^\dag b\rangle $ is the total number operator subject to losses. 
The squeezing is in general defined for an arbitrary axis, which makes it possible to optimize. However it is more practical to examine a particular axis in the lab frame, which we choose it to be the $x$-axis. 

\begin{figure}[!tbp]
\begin{center}
\includegraphics[width=8.6 cm]{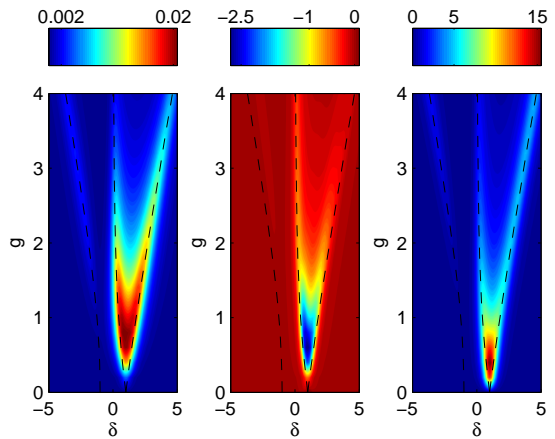}
\caption{\label{fig:entropy}(Color online) Same as Fig. \ref{fig:transmissions}, but for the von Neumann entanglement entropy $S$ (Left Figure), genuine two-mode entanglement parameter $\lambda_{ab}$ (Middle Figure), and the excited state population $\sigma_{ee}=\sigma^\dag \sigma$ (Right Figure). $\lambda_{ab}$ is multiplied by $10^5$ and $\sigma_{ee}$ is multiplied by $10^3$ for visibility. Dashed lines indicate the probe resonances with the single excitation manifold. $g/2\pi$ and $\delta/2\pi$ are in MHz units.}
\end{center}
\end{figure}

According to Fig.\ref{fig:entropy}, $S$ is strongest about $M_0=(g,\delta)\sim (J/2,J)$. For $g>J$, it decreases, but occur predominantly along the polariton branches. The decrease with $g$ is due to reduction of the Hilbert space dimensionality. All the three subsystems, two-level atom and the radiation modes, are excited About $M_0$ point which can be seen in Figs. \ref{fig:transmissions}-\ref{fig:entropy}. Three component atom-cavity "molecular" structure reduces to two-component forms as $g$ increase. The middle and right polariton branches becomes dominated by $a$ and $b$ modes, respectively. Two-level atom excitation is mainly along the right branch. The middle figure indicates genuine two-mode correlations are weakest about $M_0$ and stronger along the middle branch,. The opposite behavior of $\lambda_{ab}$ and $S$ confirms the strong influence of atomic component on $S$.  $\lambda_{ab}<0$ reflects that $a-b$ correlations are atom mediated rather than being genuine. Though not visible in the scale of the figure, $\lambda_{ab}>0$ is found in the vicinity of $(\delta=1,g=0)$ point, for which decay channels of the two-level atom is turned off and true two-mode entanglement is survived, though barely, in the steady state. 

\begin{figure}[thbp]
\begin{center}
\includegraphics[width=8.6 cm]{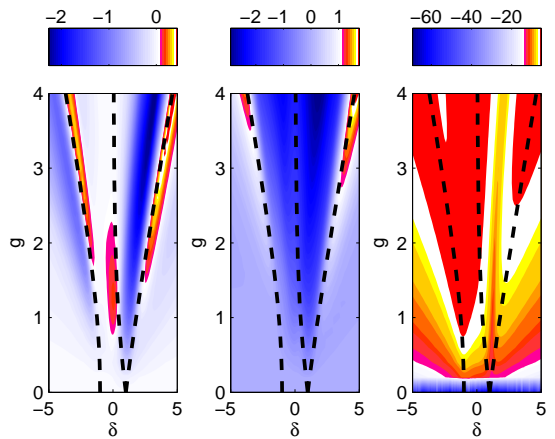}
\caption{\label{fig:g2ag2bxi}(Color online) Same as Fig. \ref{fig:transmissions}, but for second-order coherence functions $g_a^2(0)$ (Left Figure), $g_b^2(0)$ (Middle Figure), and the spin squeezing parameter $\xi^2_x$ (Right Figure). These are plotted as $-\log(g_a^2(0)),-\log(g_b^2(0))$, and $-\log(\xi_x^2)$. $g/2\pi$ and $\delta/2\pi$ are in MHz units.}
\end{center}
\end{figure}

The second order coherence functions for the modes $a,b$ together with the spin squeezing parameter of multi-partite entanglement are shown in Fig. \ref{fig:g2ag2bxi}. We plot $-\log(g_a^2(0)),-\log(g_b^2(0))$, and $-\log(\xi_x^2)$ so that their positive values respectively mean sub-Poisson statistics for modes $a,b$ and particle entanglement of optical and microwave photons. 

The behavior of the $g_b^2(0)$ for $g>J$ is typical in JC model. For $g<J$,  $g_b^2(0)\sim g_a^2(0)\lesssim 0$. The equivalence of the $g^2(t)$ to Mandel Q parameter \cite{qOpticsBook} associates anti-bunched photons with $g^2<0$ with sub-Poissonian statistics. The only significantly populated  anti-bunching region for mode $a$ is about $\delta\sim -0.5$, at the beginning of the optical domain. The other two anti-bunching regions of the optical mode are along the microwave domains and hence relatively weakly populated. The coherence properties of the microwave photon is transferred to the optical one via their common interaction with the two-level. Collective enhancement of the interaction coefficients induces bosonic, linear oscillator, character to the $\Delta$-type ensemble, which is harmful for sub-Poisson photon statistics. By addition of the two-level atom, dressed system regains its lost anharmonic character. Enhancement of both $J$ and $g$ optimizes the anti-bunching of the optical photons.
Despite relatively small coupling constants, the minimum $g_a^2(0)\sim 0.8$ and $g_b^2(0)\sim 0.1$  are within the same range with two-level nonlinear oscillators. A promising route to enhance these numbers could be to consider a more efficient source of anharmonicity, such as the $N$-type $4$-level system \cite{NtypeAtom}. 

Nonlinearity of the JCM is associated with the anharmonic spacing of the energy levels. One prominent effect of anharmonic level spacing is photon blockade \cite{photonBlockade} which is the suppression of the sequential absorption of two photons into the system.  In our case the anharmonicity influences the normal modes of the coupled $a,b$ modes similar to a case of photon blockade in a two-mode cavity.  
The degree of second order coherence is equivalent to Mandl Q parameter which is a relative measure of number fluctuations. Photon blockade is associated with sub-Poisson statistics of the anti-bunched radiation. Since the normal mode transformation is a rotation in the mode space, sub-Poisson statistics that is found in a normal mode can be translated into the original modes. Such quantum noise description of the effect allows us to appreciate the localization of the relatively strong particle correlations about the polariton branches.    

Due to strong dissipation channels and relatively weak coupling strengths the steady state exhibits low level of quantum correlations among the subsystems. On the other hand, some of its promising features can be found in early time quantum dynamics of the system. Our objective in the next section would be to go beyond steady state and explore some non-equilibrium cases for our model.
 
\subsubsection{Non-Equilibrium Coherence and Correlations} 

We investigate dynamics of quantum correlations among the radiation modes in the single ensemble case for  initial state $\ket{\psi(0)}=\ket{11g}$. Truncating the Fock space for each radiation mode at $7$ photons, quantum master equation is solved by the direct integration.

To comprehend clearly the effect of $g$ on the correlation dynamics, we first consider the case $\kappa_a,\kappa_b,\gamma,\gamma_\phi=0$, and accordingly, $E_a,b=0$. We choose $\delta_b=0$ for which the probe resonance with the upper polariton branch is at $\delta=\sqrt{J^2+g^2}$. 

\begin{figure}[!tbp]
\begin{center}
\includegraphics[width=8.6 cm]{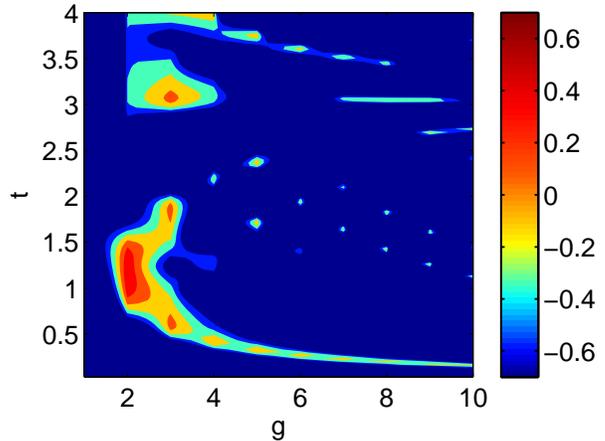}
\caption{\label{fig:noLossXi}(Color online) Time and $g$ dependence of $\xi_x^2$ when there is no
dissipation in the system for an initial state $\ket{\psi(0)}=\ket{11}_{ab}\ket{g}$ at $\delta_b=0$ and $\delta = \sqrt{g^2+J^2}$ with $J/2\pi=1$ MHz. In the figure $g/2\pi$ is in MHz and $t$ is in $\mu$s units.}
\end{center}
\end{figure}

By calculating the $\xi_x^2$, we find that there is no multi-particle entanglement for $g\le J$. We examine $\xi_x^2$ for $g> J$ in Fig. \ref{fig:noLossXi}. We plot $-\log_2{\xi_x^2}$ so that any positive value of it would be the indicator of the particle entanglement. Strongest multi-particle correlations exist
when $g\sim J$. Further increase of $g$ gives weaker squeezing. While the entanglement happens earlier at larger $g$, its duration
severely decreases with $g$. 

\begin{figure}[!tbp]
\begin{center}
\includegraphics[width=8.6 cm]{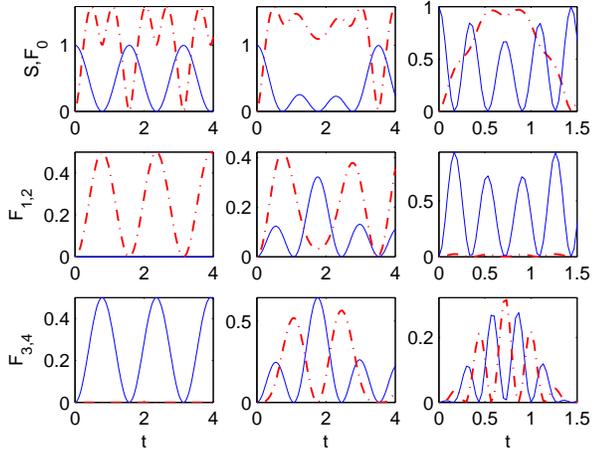}
\caption{\label{fig:fideltyNoLossBest}(Color online) Time dependence of $S$ and fidelities $F_j,j=0,1,2,3,4$ of the states $\ket{11g},\ket{10e},\ket{20g},\ket{02g},\ket{01e}$, respectively.
There is no dissipation in the system and the initial state is  $\ket{\psi(0)}=\ket{11}_{ab}\ket{g}$. $\delta_b=0$ and $\delta = \sqrt{g^2+J^2}$ with $J/2\pi=1$ MHz are used. In the figure $t$ is in $\mu$s units. The left, middle and right columns are for $g/2\pi=0,1,10$ MHz. Dash-Dotted curves in the first, second and third rows are for $S,F_2,F_4$.}
\end{center}
\end{figure}

In the regime $g<J$ lacking multi-particle entanglement, there is strong mode entanglement as depicted in the first column of Fig. \ref{fig:fideltyNoLossBest}, where we plot the $S$ together with the fidelities $F_j=Tr(\ket{\psi_j}\bra{\psi_j}\rho)$  of the states  $\ket{\psi_{j=0..4}}=\ket{11g},\ket{10e},\ket{20g},\ket{02g},\ket{01e}$, respectively. For $g\sim 0$ there are two maximally entangled two-particle states \cite{cyclic_ensemble_model}. The one appearing at the peaks $S\sim 1.5$ is $\ket{\psi}=(\ket{20g}+\ket{02g}\pm i\ket{11g})/\sqrt{3}$. It is entangled in three dimensional space so that the maximum $S=log_2(3)\sim 1.5$.  While the one at the peaks $S\sim 1$ is of the form $\ket{\psi}=(\ket{20g}+\ket{02g})/\sqrt{2}$ and is entangled in two dimensional state space.  At regular times the system is disentangled to $\ket{\psi_0}$. 
Strong and long-lived mode entanglement can be found when $g\sim J$ as shown in second column of the Fig. \ref{fig:fideltyNoLossBest}. Splitting of the degenerate states helps to preserve three dimensional structure of the subspace of entanglement to ensure highest available $S\sim 1.5$. In addition, destruction of the disentanglement ensures high $S$ can be maintained relatively longer duration in comparison to $g\sim 0$ regime. For $g\gg J$, according to the last column of Fig. \ref{fig:fideltyNoLossBest}, dynamical behavior of $S$ resembles qualitatively $g\sim 0$ regime, but for the lower $S\sim 1$ values. The states at these points are constructed by $\ket{11g}$ and $\ket{01e}$ for the middle, lower peak, and by
$\ket{10e}$ and $\ket{02g}$ for the higher peaks. They are both in two dimensional state spaces so that $S\sim 1$.

\begin{figure}[!tbp]
\begin{center}
\includegraphics[width=8.6 cm]{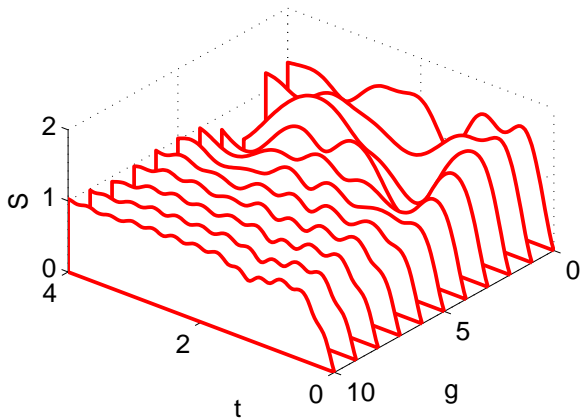}
\caption{\label{fig:gooddissipationS}(Color online) Time and $g$ dependence of von Neumann entropy for an initial state $\ket{\psi(0)}=\ket{11}_{ab}\ket{g}$ at $\delta=\sqrt{g^2+J^2}$ and $\delta_b = 0$. In the figure $g/2\pi$ is in MHz and $t$ is in $\mu$s units. The decay rates are $\kappa_a/2\pi,\kappa_b/2\pi,\gamma_\phi/2\pi=0.1$ MHz. }
\end{center}
\end{figure}

\begin{figure}[!tbp]
\begin{center}
\includegraphics[width=8.6 cm]{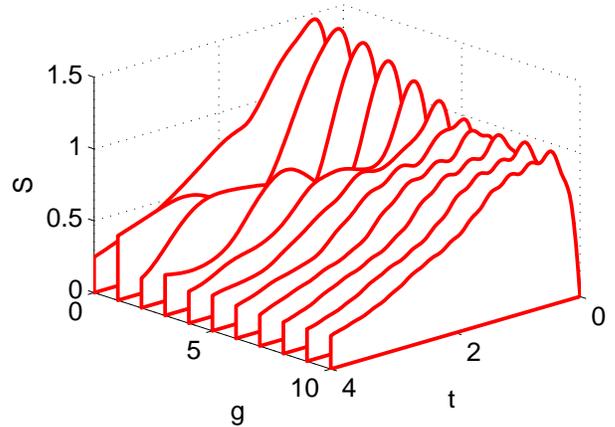}
\caption{\label{fig:dissipationS}(Color online) Same as Fig. \ref{fig:gooddissipationS}, but for $\kappa_a/2\pi=0.4,\kappa_b/2\pi=0.4,\gamma_\phi/2\pi=0.3$ MHz and
$\delta_b=J$.}
\end{center}
\end{figure}

We now examine the behavior of correlations in the presence of dissipation.
A typical result for our simulations is shown in Fig. \ref{fig:gooddissipationS}
indicating the dynamical behavior of $S$ for the same parameters with the lossless case but for $\kappa_a/2\pi,\kappa_b/2\pi,\gamma_\phi/2\pi=0.1$ MHz. If we use the larger decay rates, same with those used in the steady state analysis, and take the same  $\delta_b=J$ with the steady state case, 
we find the result in Fig. \ref{fig:dissipationS}. We see that the features of mode entanglement discussed in dissipationless case is vastly available in the case of moderate damping. In the larger damping case, except the early times, entanglement cannot be obtained. Some signatures of the prolonged entanglement duration is still found in the $g\sim J$ even under such strong damping condition. 

\begin{figure}[!tbp]
\begin{center}
\includegraphics[width=8.6 cm]{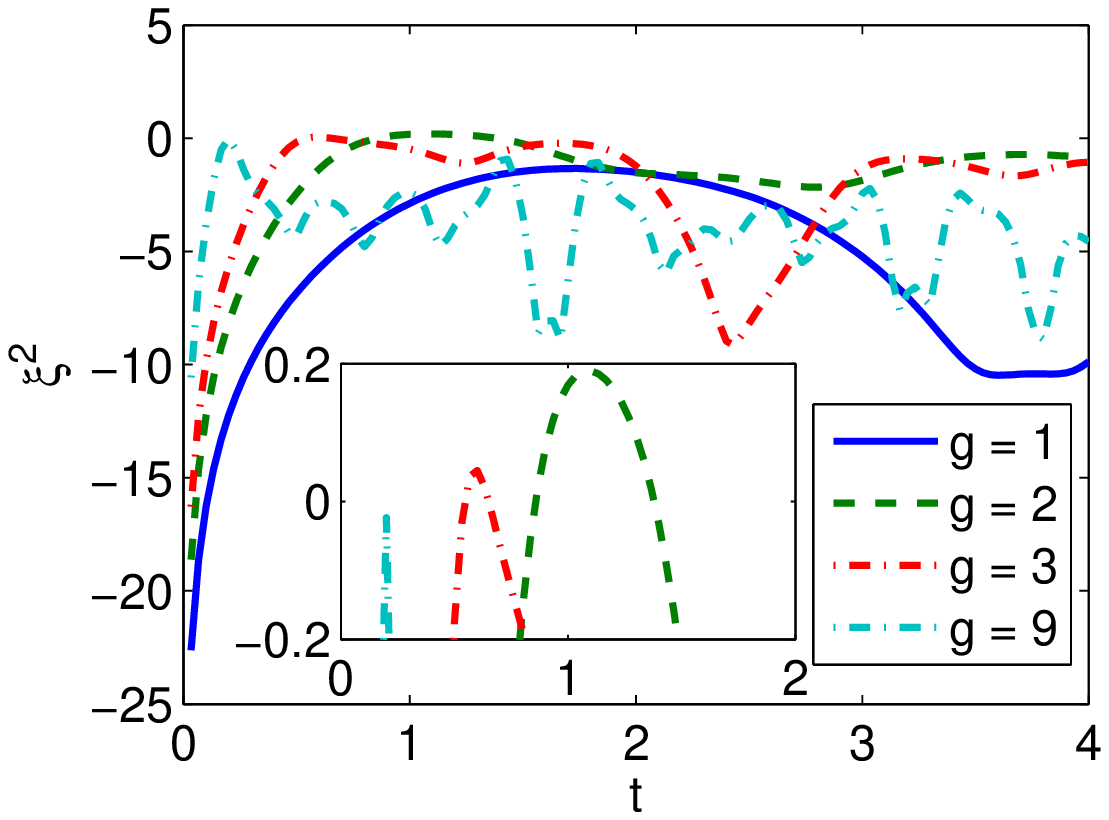}
\caption{\label{fig:gooddissipationXi}(Color online) Same as Fig. \ref{fig:gooddissipationS} but for the squeezing parameter $\xi^2_x$.}
\end{center}
\end{figure}

\begin{figure}[!tbp]
\begin{center}
\includegraphics[width=8.6 cm]{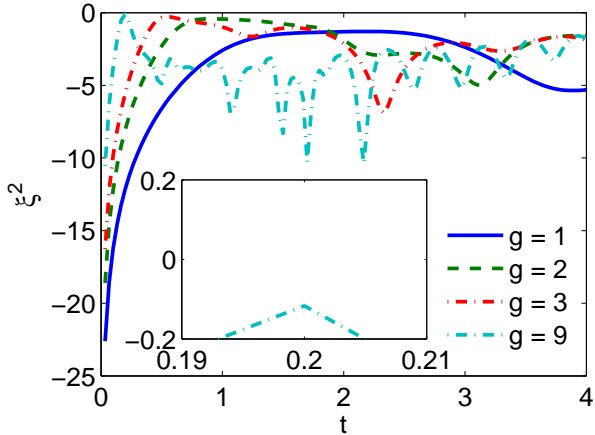}
\caption{\label{fig:dissipationXi}(Color online) Same as Fig. \ref{fig:dissipationS} but for the squeezing parameter $\xi^2_x$}
\end{center}
\end{figure}

Figs. \ref{fig:gooddissipationXi}-\ref{fig:dissipationXi}, show that multi-particle entanglement can still be found at certain $g$ values at particular time domains for weaker damping situation, while no spin squeezing and entanglement remain in the larger dissipation case.

%
%
%

\subsection{Two Atomic Ensembles}

We shall now consider the case of two identical, symmetrically placed cyclic ensembles. In contrast to $a-b$ coupling, there is no direct $a_1-a_2$ interaction now. We did not find multi-partite entanglement between the photons of $a_1$ and $a_2$ modes. We focus on mode entanglement and population dynamics in this case. Since the system has three mode radiation modes, we take into account possibility of genuine three-mode entanglement. We use an extended set of mode correlation parameters \cite{GenuineModeEnt}
\begin{eqnarray}
\lambda_{1b}&=&\mid \langle a_1^\dag b\rangle\mid^2-\langle a_1^\dag a_1b^\dag b\rangle,\\
\lambda_{2b}&=&\mid \langle a_2^\dag b\rangle\mid^2-\langle a_2^\dag a_2b^\dag b\rangle,\\
\lambda_{12}&=&\mid \langle a_1^\dag a_2\rangle\mid^2-\langle a_1^\dag a_1a_2^\dag a_2\rangle.
\end{eqnarray}
When any two of them become simultaneously positive, genuine three-mode entanglement is found. Typical example is the so called three parity W state \cite{Wstate} which is defined to be $W=(\ket{100}+\ket{010}+\ket{001})/\sqrt{3}$. In our case the Hilbert space is enlarged by the two-level atom. We denote basis states with
$\ket{n_1,n_b,n_2,s}$ where $n_{1,2},n_b$ are the occupation numbers of the radiation modes $a_{1,2},b$ and $s= e,g$.

We solve the density matrix equations using quantum trajectory method. We take $25$ trajectories and truncate Fock space at $2$ photons in each radiation mode $a_{1,2},b$. We fix $\delta_b=0$ and 
$\delta=\sqrt{J^2+g^2}$ for a given $g$ for which the probes are in resonant with the upper polariton branch of the single excitation manifold. For arbitrary $\Delta$ there are four such branches now due to additional optical degree of freedom in comparison to the single-ensemble case with three polariton branches. We explore dynamical evolution of two initial preparations of the system: (i) $\ket{\psi(0)}=\ket{010g}$ and (ii)  $\ket{\psi(0)}=\ket{100g}$.

\begin{figure}[!tbp]
\begin{center}
\includegraphics[width=8.6 cm]{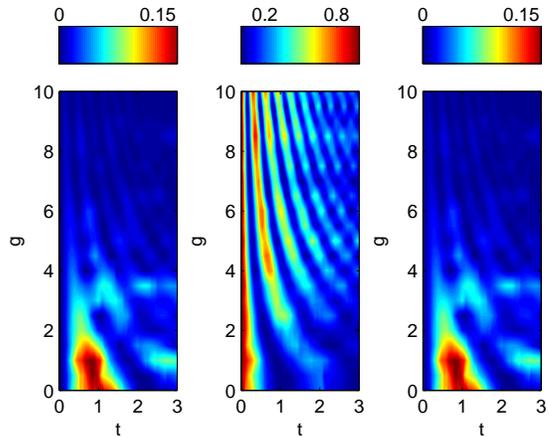}
\caption{\label{fig:localizeVSsplitCavityPhoton}(Color online) Time dependence of populations of the radiation modes $n_1=\langle a_1^\dag a_1\rangle$ (Left Figure), $n_b=\langle b^\dag b\rangle$ (Middle Figure), $n_2=\langle a_2^\dag a_2\rangle$ (Right Figure) for an initial state $\mid\psi(0)=\ket{010g}$. We take $J_{1,2}/2\pi=1$ MHz, $\delta_b=0$,$\delta=\sqrt{g^2+J^2}$, $\kappa_{1,2,b}=(2\pi)0.4$ MHz,
$\gamma/2\pi=0.02$ MHz, $\gamma_\phi/2\pi=0.3$ MHz, $E_{x}=0.1\kappa_x, x=1,2,b$. }
\end{center}
\end{figure}

In the first case, our simulations reveal that  when $g<J$ population of the microwave cavity photon is symmetrically split into to the optical modes as shown in Fig. \ref{fig:localizeVSsplitCavityPhoton}. The effect is an optical analog of coherent population trapping \cite{CPT}. At large $g>J$ splitting is suppressed and the cavity photon is predominantly making localized interactions with the two-level atom. To comprehend this effect let us write the dark state $\ket{DS}$ in the single excitation manifold for the two-ensemble model as follows
\begin{eqnarray}
\ket{DS}=\frac{1}{\sqrt{J^2+g^2}}\left(-J\sigma^\dag-\frac{g}{\sqrt{2}}a_1^\dag-
\frac{g}{\sqrt{2}}a_2^\dag\right)\ket{vac},
\end{eqnarray} 
with $\ket{vac}=\ket{000g}$ being the vacuum state of the composite system. For $g\gg J$, $\ket{DS}$ becomes $\ket{DS}\approx - (\ket{100g}+\ket{001g})/\sqrt{2}$, while for $g\ll J$ it gets $\ket{DS}\approx \ket{000e}$. This suggests that in the regimes where the long range interaction $J$ dominates over the short range interaction $g$, the dark state $\ket{000e}$ is inaccessible by the initial state $\ket{\psi(0)}=\ket{010g}$. As such, $\ket{\psi(0)}$ only yields a coherent transfer of $b$ population to the $a_{1,2}$ modes. According to Fig. \ref{fig:localizeVSsplitCavityPhoton} population transfer is complete in about $t\sim 1\,\mu$s. Population exchange is a coherent process happening at regular time intervals of every $2J$. At lower damping rates, such as $\gamma,\kappa_{1,2,b}/2\pi \sim 0.1$ MHz, we find a few more cycles of transfer. 

\begin{figure}[!tbp]
\begin{center}
\includegraphics[width=8.6 cm]{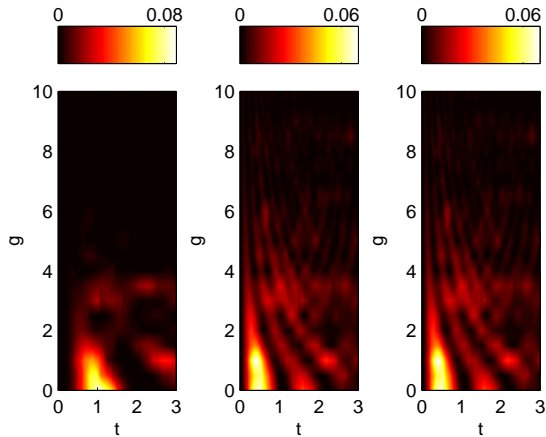}
\caption{\label{fig:3modeEntanglementSplittedCavity}(Color online) Same as Fig. \ref{fig:localizeVSsplitCavityPhoton} but for the genuine mode entanglement parameters $\lambda_{12}$ (Left Figure),  $\lambda_{2b}$ (Middle Figure), and  $\lambda_{1b}$ (Right Figure). }
\end{center}
\end{figure}

When the tunneling from $b$ to $a_{1,2}$ is complete, the system radiates genuine two-mode entangled optical photons which can be verified by the Fig. \ref{fig:3modeEntanglementSplittedCavity}. We verified $S\sim 1$ is a reliable bi-partite entanglement signature at the corresponding times for low ($\gamma,\kappa_{1,2,b}/2\pi \sim 0.1$ MHz) damping. During the course of transfer, at about $t\sim 0.5\,\mu$s all three modes are populated. At such times, as the excited state of the two-level atom remains in the dark, three parity $W$ state with genuine three-mode entanglement is found. Corresponding $S\sim 0.7$ is relatively lower than the maximally entangled two-mode state. Due to its robustness against single particle losses relative to 
Greenberger-Horne-Zeilinger
(GHZ) states \cite{GHZstate1,GHZstate2} ($\ket{GHZ}=(\ket{000}+\ket{111})/\sqrt{2}$)
and its potential role in quantum networks, $W$ states have appealing features in quantum information science. They are proposed for bi-partite entanglement distillation \cite{distillEPRfromW}. Our results provide a promising setting to realize these goals. 

In the $g\gg J$ case, $\ket{DS}\approx -(\ket{100g}+\ket{001g})/\sqrt{2}$ remains in the dark. The initial state is predominantly coupled to $\ket{000e}$.
The radiation energy is trapped in the microwave cavity ($b$) mode and exchanged locally with the two-level atom. Associated Rabi oscillations are shown in the middle figure of Fig. \ref{fig:localizeVSsplitCavityPhoton}. Mode entanglement parameters practically vanish in this case, as shown in Fig. \ref{fig:3modeEntanglementSplittedCavity}. 

\begin{figure}[!tbp]
\begin{center}
\includegraphics[width=8.6 cm]{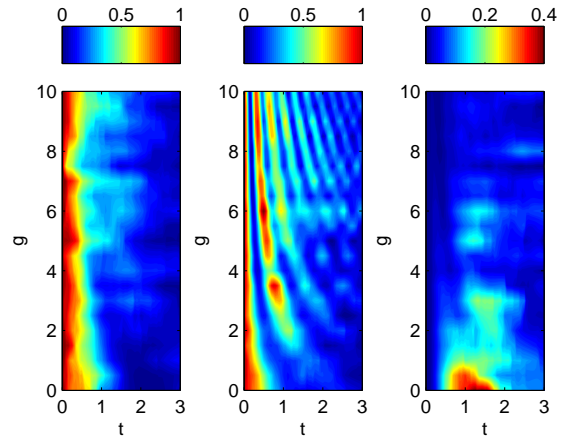}
\caption{\label{fig:blockMode_aByCavity2}(Color online) Same as Fig. \ref{fig:localizeVSsplitCavityPhoton}, but for an initial state $\mid\psi(0)=\mid 110\rangle\mid g\rangle$. }
\end{center}
\end{figure}

We consider an asymmetric populated initial state to show  that instead of two-mode entanglement of optical modes, optical and cavity mode entanglement can be emphasized. For this aim we start with $\ket{\psi(0)}=\ket{110g}$. Population dynamics at various $g$ is depicted in Fig. \ref{fig:blockMode_aByCavity2}. 

\begin{figure}[!tbp]
\begin{center}
\includegraphics[width=8.6 cm]{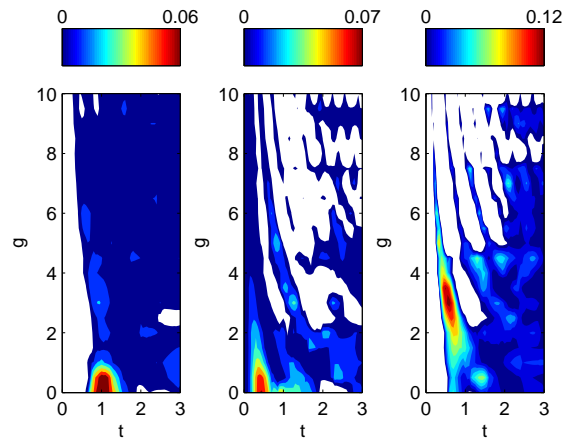}
\caption{\label{fig:MirrorCaseGenuine3modeEntanglement}(Color online) Three-mode correlation parameters that can be used to reveal Genuine three-mode entanglement in the evolution of an initial state $\mid\psi(0)=\mid 110\rangle\mid g\rangle$. }
\end{center}
\end{figure}

This localization of the cavity photon in the $g>J$ Rabi oscillations regime can be used to block an optical mode to access the other distant optical mode and as such, the corresponding distant ensemble. The localizations of the radiation energies in $g\gg J$ regime can be seen as blocking of an optical mode interacting with an ensemble to access a distant ensemble and the corresponding optical mode via the microwave nonlinear cavity. In $g\ll J$ limit, the $b$ mode still splits into two but its tunneling into $a_1$ is balanced with the tunneling of $a_1$ into $b$. There is no net population current between the $a_1$ and $b$ modes, so that we see no change of population of $a_1$ mode in the left figure in Fig. \ref{fig:blockMode_aByCavity2}, except its usual decay.
Accordingly, $\lambda_{12}<0$ and $\lambda_{1b}<0$ as indicated by the white regions about $t\sim 0.5\,\mu$s in the left and the right figures of Fig. \ref{fig:MirrorCaseGenuine3modeEntanglement}. While the middle figure demonstrates that there is genuine two-mode entanglement between the $a_2$ and $b$ modes at this time. In contrast to previous $W$ state arising at this time with three-mode entanglement, initial imbalance of the populations in the present case effectively removes one particle from the $W$ state to distill or concentrate a two-mode entanglement out of it. At $t=1\,\mu$s the situation is similar to previous symmetric initial state and we see two-mode entanglement between the optical $a_1$ and $a_2$ modes. The effect can be viewed as the presence of dynamically distinct and transferable bi-partite entanglement phases in the multi-mode system.
%
%
While we consider here only the simplest symmetric and identical two-ensembles time dependent controllable different interaction coefficients $J_1,J_2,g$ allows for more rich coherent information and population transfer protocols.
\section{Proposal of Experimental Implementation}
\label{sec:expProposal}

\begin{figure}[!tbp]
\begin{center}
\includegraphics[width=8.6 cm]{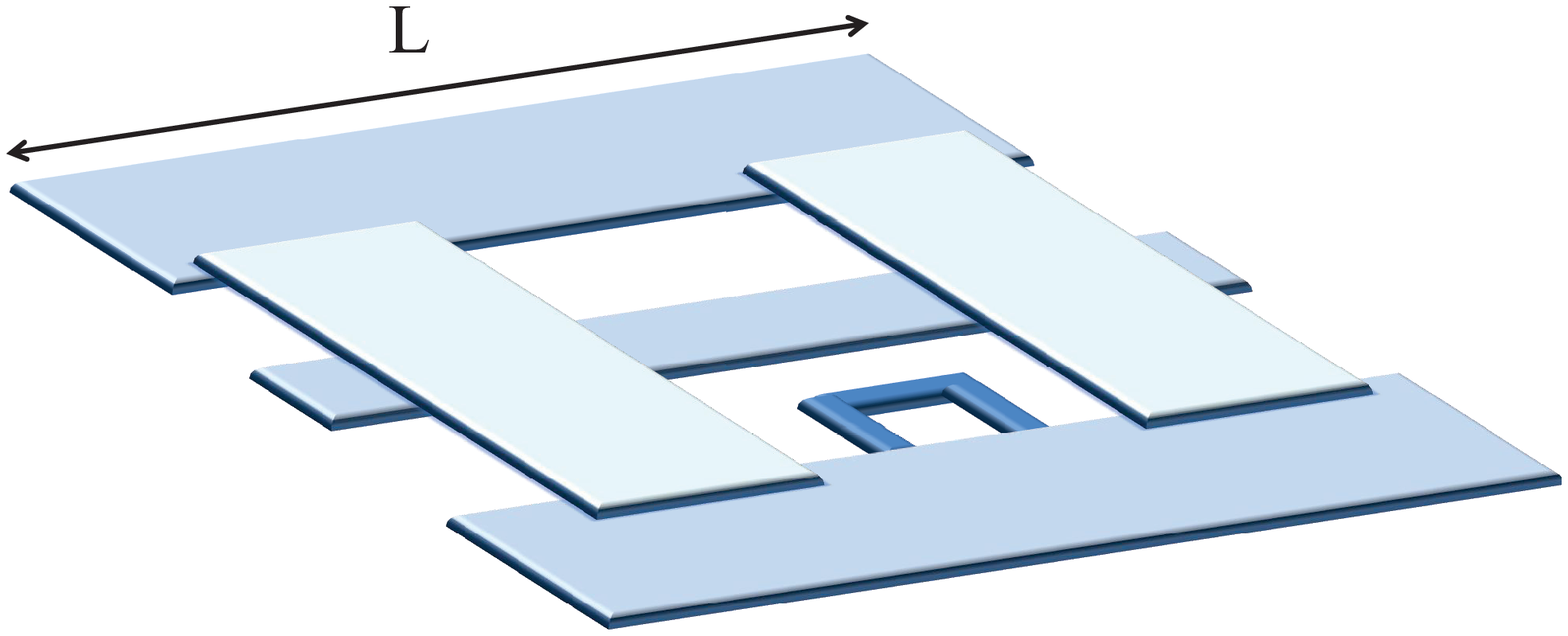}
\caption{\label{fig:proposal}(Color online) Transmon qubit and two diamond crystals couple to microwave stripline cavity of lenght $L$ and width $d$.}
\end{center}
\end{figure}

In order to realize the model Hamiltonians discussed in the preceding sections,
now we shall consider a straightforward extension of a recent experiment demonstrating strong coupling of an N-V center ensemble to a coplanar waveguide resonator \cite{NV_CPWG_exp}. The setting is shown in Fig. \ref{fig:proposal}, where two diamond crystals and a transmon qubit \cite{transmon,transmonTheory,transmonExp} strongly coupled to microwave stripline cavity \cite{cpwg_exp1}.
Diamond crystals containing color center defects serve as the quantum memory, transmon qubit serves as the quantum hardware for rapid quantum information processing, and the microwave cavity photons serve as the local data bus. By interfacing this device with external optical fields, such stationary quantum information unit would have access to optical flying qubits for quantum networking and communication at free space communication wavelengths.  

Strong coupling of a single diamond with Nitrogen-Vacancy (N-V) centers to a superconducting coplanar waveguide (CPWG) resonator has been shown recently \cite{NV_CPWG_exp}. We propose a a straight forward extension
of the physical system of Ref. \cite{NV_CPWG_exp} to the case of two diamond crystals. In order to place the diamond crystals to locations where the magnetic field is strong and homogeneous, and the transmon qubit to a similar location but for the electric field, a $L=\lambda$ cavity is considered suitable for our purpose. $\lambda\sim 50\,$mm is the microwave wavelength for the fundamental frequency of the cavity $\omega_1/2\pi\sim 6\,$GHz. The crystals are assumed at the same size of $3\times 3\times 0.5\,$ mm$^3$. Canonically quantized magnetic field of the fundamental quasi-TEM mode of the cavity is given in the lab (cavity) frame by
\begin{eqnarray}
\vec{B}(x,y,z)=\vec{B}_\perp(x,y)(b+b^\dag)\cos{\frac{\pi z}{L}},
\end{eqnarray}
where $\vec{B}_\perp(x,y)$ is the transverse field distribution.
Placement of the crystals $\lambda/2$ apart, at the magnetic field antinodes of the fundamental mode along the cavity axis ($x$), allows for sufficiently strong magnetic coupling to the cavity mode as well as adequate space for optical access to each crystal. At the same time, transmon qubit placed in the middle of the crystals would be at the electric field antinode of the fundamental cavity mode. It is now well-established that, under certain conditions on the Cooper-pair box and circuit QED system, Jaynes-Cummings model can describe the
coupling of the qubit to the single mode cavity \cite{CooperPairBoxJCM,transmon}. 

In order to determine the conditions to obtain the model Hamiltonians we have used in the preceding sections, thus we shall focus here the relatively less explored N-V center coupling to the cavity, in particular we examine if such a coupling can be described by the $Delta$-type transition model. 

\begin{figure}[!tbp]
\begin{center}
\includegraphics[width=8.6 cm]{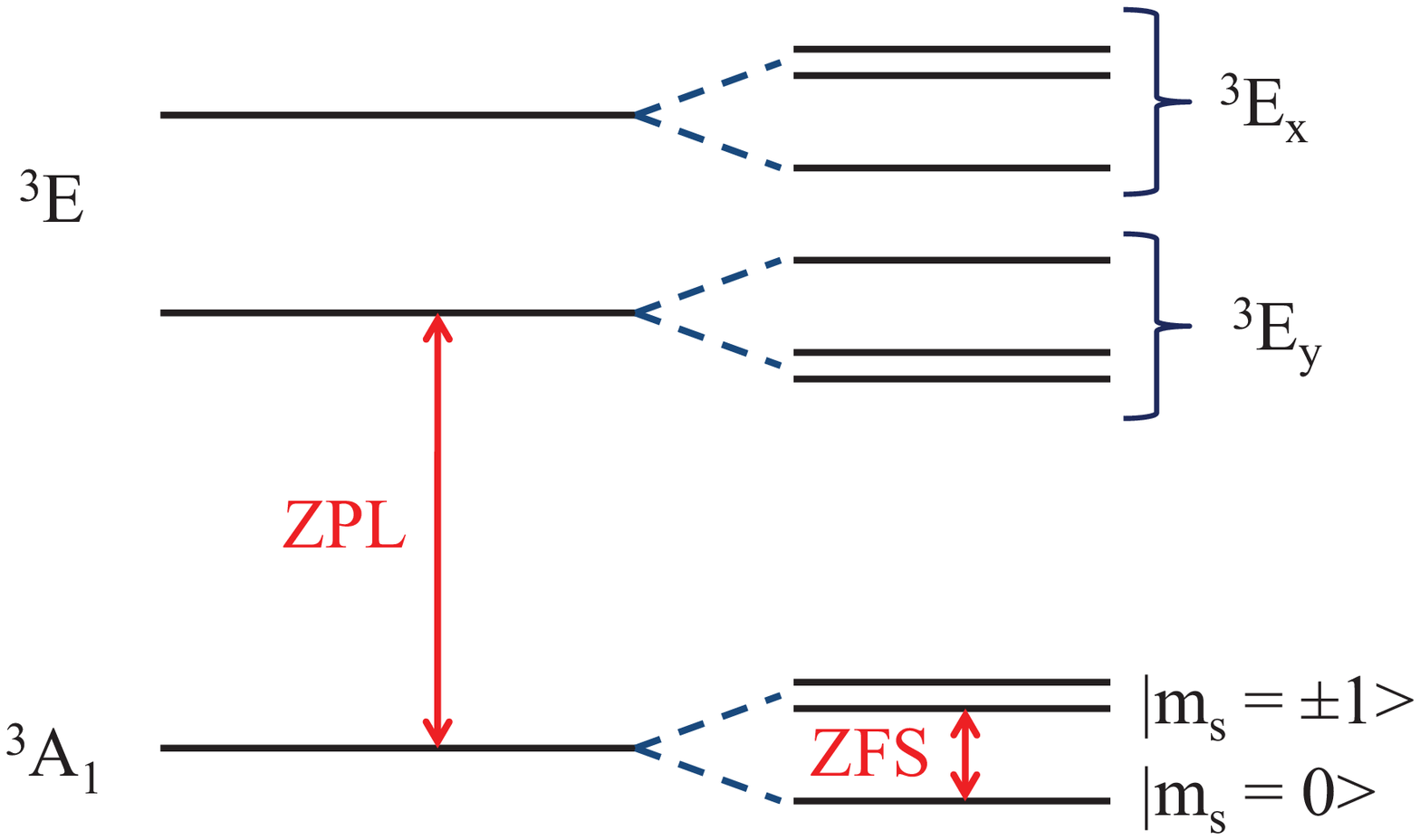}
\caption{\label{fig:NV_Levels}(Color online) Energy level diagram of negatively charged N-V center. ZPL is the zero
phonon line at $638$ nm ($1.945$ eV) and ZFS is the zero field splitting at
$2.87$ GHz.}
\end{center}
\end{figure}

The diagram shown in Fig. \ref{fig:NV_Levels} describe the relevant energy levels of a negatively charged N-V center \cite{NVLevelsTheory,NVLevelsExp}. The ground state manifold is $^3A$, with zero field splitting $D\sim 2.87$ GHz, the excited state is $^3E$ separated from the ground level by the zero phonon line at $637$ nm, and there is a metastable state $^1A$ with non-radiative transitions. Further splittings of $^3E$ level are in the order of $2.6$ GHz and $2.3$ GHz for $^3E_x$ and $^3E_y$, respectively.     

Electric dipole (E1) and magnetic dipole (M1) selection rules forbid an immediate realization of Raman type ($\Lambda$) transitions in the given energy level diagram, let alone a $\Delta$ transition. On the other hand, optical Raman induced spin-flipping processes \cite{spinflipExp1,SpinFlipExp2,spinflipNVCexp} have been realized in addition to spin conserving (cyclic) transitions \cite{spinPreservingExp} in recent experiments. To comprehend how this happens, detailed recent studies of the excited level reveal that among the potential spin-spin and spin-orbit interactions that can influence the excited level structure, it is the local strain induced level mixing that allows for the essential non-spin preserving transitions for an optical $\Lambda$-scheme. Using external magnetic or electric fields or specifically engineered N-V centers thus one can produce the level combinations for such transitions. To make our discussion concrete, we generalize the $\Lambda$-transition generation protocol using external electric fields \cite{lambdaNVbyE} to the case of $\Delta$-transition. Typical electric field strengths several $\sim MV/m$ are experimentally used \cite{spinflipNVCexp}.

\begin{figure}[!tbp]
\begin{center}
\includegraphics[width=8.6cm]{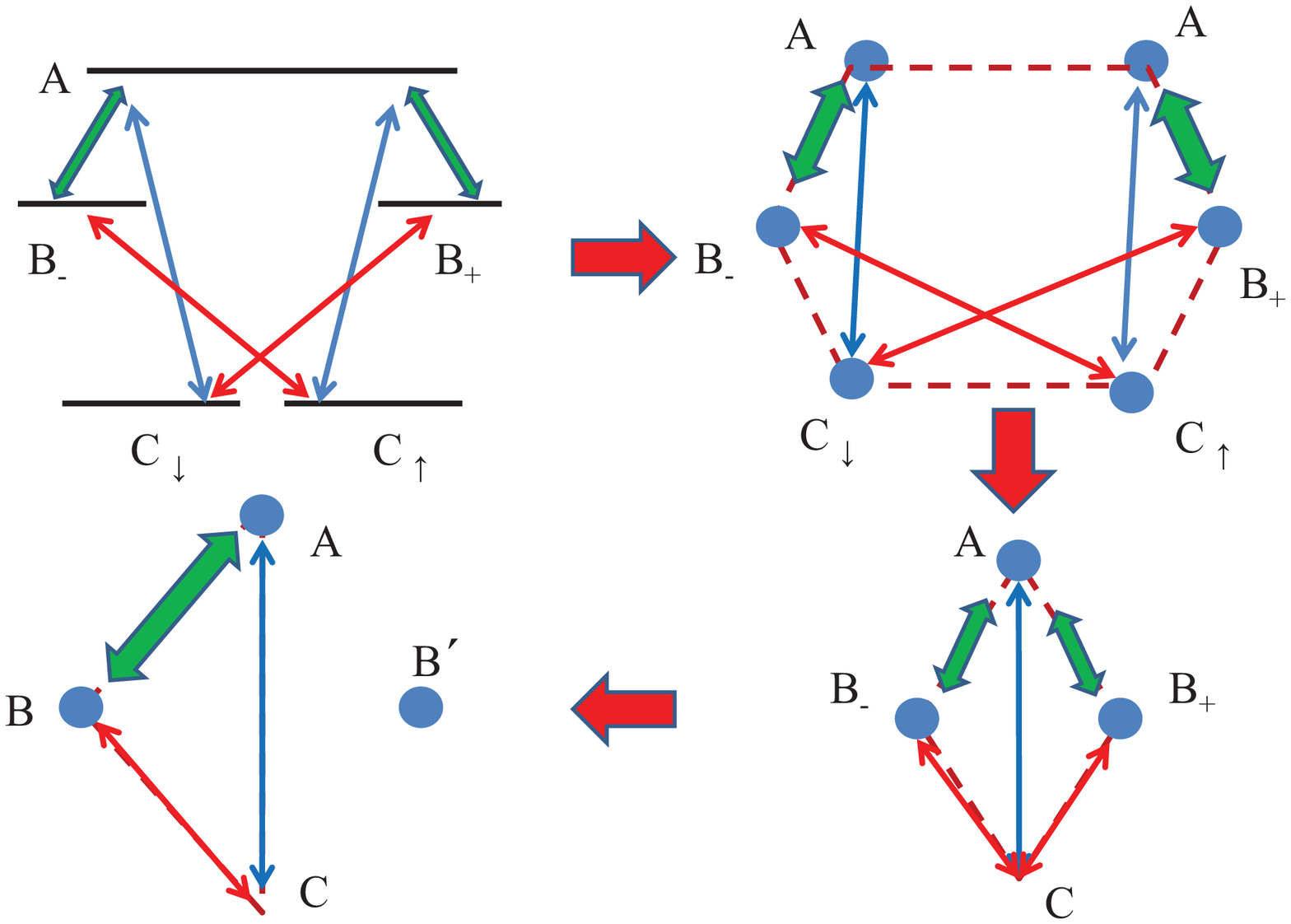}
\caption{\label{fig:mStransformedLvs}(Color online) Energy level diagram of negatively charged N-V center. ZPL is the zero
phonon line at $638$ nm ($1.945$ eV) and ZFS is the zero field splitting at
$2.87$ GHz.}
\end{center}
\end{figure}

Under strong electric field the level structure of the negatively charged N-V center is shown in the Fig. \ref{fig:mStransformedLvs}. Here, both cyclic (spin preserving) and spin-flipping transitions can simultaneously happen. 

Optical fields are right circular polarized. E1 and ESR selection rules allows us to obtain cyclic transitions in this single N-V molecule in a hexagon scheme, where all nodes can be accessed starting from any other node, as shown in Fig. \ref{fig:mStransformedLvs} (b). The upper level $A$ is an excited level generated by the mixed states by the electric field. The intermediate levels are $\ket{m_s=-1,m_I=\uparrow}$ and $\ket{m_s=+1,m_I=\downarrow}$, denoted by $\ket{-}$ and $\ket{+}$, respectively. The lowest levels are $\ket{m_s=0,m_I=\uparrow}$ and $\ket{m_s=0,m_I=\downarrow}$, denoted by $\ket{\uparrow}$ and $\ket{\downarrow}$, respectively. Coupling of the cavity field to the ground state $^3A$ manifold is described by an anomalous Zeeman interaction of the form
\begin{eqnarray}
H=\sum_i\sigma_{i}^T\tilde{D}\sigma_{i}+\mu_B\sum_i\vec{B}_i^T\tilde{g}\vec{\sigma}_i,
\end{eqnarray}
where $\mu_B$ is the electronic Bohr magneton and $\tilde{g}$ is a diagonal tensor with elements $(g_\perp,g_\perp, g_\parallel)$ with 
$g_\perp=2.0024$ and $g_\parallel = 2.0028$ in a frame where the quantization axis ($z$) is chosen to be the symmetry axes of the N-V molecule. In the $C_{3v}$ symmetry of diamond, one can find four of such axes corresponding to four crystallographic classes $NV-f, f=1,2,3,4$. We work a within a particular class, but omit its label ($f$) for now for the sake of notational simplicity. Expressions for different classes can be obtained by using spherical tensor operations. Nuclear Zeeman term is too weak and ignored. Depending on the relative density of surrounding nuclei ($^14$N,$^15$N,$^13$C), quadrupole and hyperfine interactions can impose significant decoherence on the central electronic spin \cite{NV_CPWG_exp}. Their contribution can be attributed to our effective loss terms in the master equation treatment we used. The first term is the zero-field Hamiltonian, with $\tilde{D}$ being the zero-field splitting tensor, which is diagonal in the molecular frame with the elements $(D_\perp,D_\perp,D_\parallel)$. It simplifies to $H_{ZFS}=D\sigma_z^2$, up to a constant, with $D=3D_\parallel/2\sim 2.87$ GHz. The index $i=1,...,N_f$ is the index of N-V defects at location $\vec{r}_i=(x_i,y_i,z_i)$ in the lab (cavity) frame. We shall consider a single crystal case first. 

Writing the cavity field as $\vec{B}(\vec{r}_i)=\vec{{\cal B}}_i(b+b^\dag)$, with
$\vec{{\cal B}}_i=\vec{B}_\perp(x_i,y_i)\cos{(\pi z_i/\lambda)}$, the Hamiltonian becomes in the ground state basis
\begin{eqnarray}
H=D\sum_{i=\pm}X_{ii}+g_B(b+b^\dag)(X_{\downarrow +}+X_{\uparrow -} +h.c.),
\end{eqnarray}    
where we neglected the small Zeeman shift, which is in the order of kHz for a cavity field in the order of nT, next to the $D \sim$ GHz. Generalized Hubbard operators are introduced to be
\begin{eqnarray}
X_{\downarrow +}&=&\sum_i\frac{{\cal B}_{i+}}{\cal{B}_\perp}R_{\downarrow +}^{(i)},\\
X_{\uparrow -}&=&\sum_i\frac{{\cal B}_{i-}}{\cal{B}_\perp}R_{\uparrow +}^{(i)},\\
X_{+-}&=&\sum_i R_{+-}^{(i)}.
\end{eqnarray}
Here, ${\cal B}_{i\pm}={\cal B}_{ix}\pm{\cal B}_{ix}$ and 
\begin{eqnarray}
{\cal{B}_\perp}=\sqrt{\sum_i\left( {\cal B}_{ix}^2+ {\cal B}_{iy}^2
\right)}.
\end{eqnarray}

In the large $N_f$ and weak excitation limit, for a quasi-uniform magnetic field distribution, we can employ the group contraction method as before, to get a bosonic representation for the Hubbard operators such that 
$X_{\downarrow +}\sim B_+$ and $X_{\uparrow -}\sim B_-$. Consistently, they form the Schwinger representation of the remaining isospin subgroup so that we take
$X_{+-}\sim B_+^\dag B_-$ which allows for replacing $X_{++}+X_{--}$ with 
$B_+^\dag B_+ + B_-^\dag B_-$ in the zero field splitting term of the Hamiltonian. 

The interaction of the excited state with the optical fields can be described in terms of the collective operators as well. To make this consistently with the microwave transitions we first make the usual bosonization of ground level to excited level $\mid A\rangle$ transition via $X_{\uparrow A}\sim X_{\downarrow A}\sim A$. Here we implicitly assume the populations in degenerate ground levels are large and $A$-mode is weakly excited. Then consistency with the isospin groups is satisfied by taking their Schwinger representations as 
$X_{A\pm}=A^\dag B_\pm$.
To achieve this limit adequately, we demand that the optical fields are to be made sufficiently homogeneous and isotropic over the diamond substances. This may be accomplished by suitable confocal microscopy. The total Hamiltonian in this bosonic limit describes again a cyclic transition system which is reduced from a hexagon to a diamond scheme, as shown in Fig. \ref{fig:mStransformedLvs} (c). The degeneracy of intermediate levels allows us to employ a simple multilevel Morris-Shore transformation \cite{morrisShore1,morrisShore2,morrisShore3} in the form $B=(B_-+B_+)/\sqrt{2}$ and $B^{\prime}=(B_--B_+)/\sqrt{2}$. Geometrically, this folds the diamond scheme into a delta-type cyclic transition as shown in Fig. \ref{fig:mStransformedLvs} (d) where $B^{\prime}$ is uncoupled from the system.

Even though we have described how to get a cyclic N-V ensemble, our analysis so far is limited to a single crystallographic class. To complete our discussion let us now address the case of all N-V classes. For that aim we make use of slightly more complicated Morris-Shore transformations. 

The Hamiltonian including all the N-V classes is
\begin{eqnarray}
H&=&\varpi_A\sum_{f=1}^4A_f^\dag A_f+\omega_B\sum_{f=1}^4B_f^\dag B+
\omega_a a^\dag a +\omega_b b^\dag b\nonumber\\
&&+\frac{\omega_0}{2}\sigma_z+g(b^\dag\sigma+h.c.)+\Omega\sum_{f=1}^4
(A_f^\dag B_f+h.c.)\nonumber\\
&&+\sum_{f=1}^4(g_{Bf}bB_f^\dag+g_{Af}aA_f^\dag+h.c.)
\end{eqnarray}

We find that under the conditions
\begin{eqnarray}
\frac{g_{B2}}{g_{B1}}=\frac{g_{A2}}{g_{A1}}, \quad
\frac{g_{B3}}{g_{B1}}=\frac{g_{A3}}{g_{A1}}, \quad
\frac{g_{B4}}{g_{B1}}=\frac{g_{A4}}{g_{A1}},
\end{eqnarray}
only two quasi-spin wave modes can be coupled to the electromagnetic modes. These so-called "bright" modes are given by collective bosonic modes
\begin{eqnarray}
B=\sum_{f=1}^4\frac{g_{Bf}}{g_B}B_f \quad
A=\sum_{f=1}^4\frac{g_{Af}}{g_A}A_f,
\end{eqnarray}
where
\begin{eqnarray}
g_A=\sqrt{\sum_{f=1}^4g_{Af}^2},\quad g_B=\sqrt{\sum_{f=1}^4g_{Bf}^2}.
\end{eqnarray}
The uncoupled, "dark" modes are listed in the appendix B. 
After that we arrive our starting Hamiltonians, Eqs. \ref{eq:bosonModelH0_rf}
-\ref{eq:bosonModelH1_rf} in the Sec. \ref{sec:model_sys}. A simple way to realize the dark quasi-spin wave modes is to make the interaction coefficients $g_{A/Bf}$ equal for each N-V class. This is the case in the experiment of strong coupling of the N-V center and superconducting cavity. By choosing a particular placement of the diamond crystal over the cavity substrate for which the cavity magnetic field makes effectively equal angles for the four quantization axes of the N-V classes, the Zeeman interaction coefficients become equal \cite{NV_CPWG_exp}. The collective enhancement of the interactions would depend on $N_f$, which we may assume $N_f=N/4$. Finally, the case of two diamond crystals can be immediately generalized from these results, where only the magnetic field would change a sign at the location of the other crystal half  wavelength away. 

The typical parameters in these models are, according to recent experiments as discussed in Ref. \cite{striplineCavityLoss}, as follows.
Qubit decoherence times are in the order of few $\mu$s, in particular for transmon is about $\sim 4\,\mu$s \cite{transmon}. For $T_1\sim 7\,\mu$s and $T_2\sim 500$ ns, relaxation and dephasing rates are respectively $\gamma_1/2\pi\sim 0.02$ Mhz and $\gamma_\phi/2\pi\sim 0.31$ MHz and the resonance frequency of the qubit is tunable in the ranges $\omega_0 \sim 5-15$ GHz. Stripline cavity loss rate is $\kappa/2\pi=\omega_b/2\pi Q\sim 5$ kHz  at $\omega_b/2\pi=5$ GHz in a cavity with $Q$ factor $Q=10^6$. Cavity frequency is adjustable in the ranges usually about $\omega_b/2\pi=5-10$ GHz. Coupling strength between the cavity and the qubit can be adjusted in the ranges $g \sim 5-200$ MHz. In our numerical simulations we choose our parameters ranges compatible within these practical values. In particular we fix $g_a/\Omega=0.1$ so that $J=-0.1g_b, E_a=0.1\Omega_A$. For $g_b=(2\pi)11$ MHz of Ref. \cite{NV_CPWG_exp}, we take $J/2\pi=1$ MHz. We assume $\Omega_A$ and detunings are adjusted to make $E_a=E_b$. $E_b=\sqrt{N_{cav}}\kappa_b$ with $N_{cav}=0.01$ is cavity photon number for the empty microwave cavity on probe resonance under weak drive conditions. We use $\gamma\equiv\gamma_1/2\pi=0.02$ MHz and $\gamma/2\pi=0.3$ Mhz. We consider ranges of $g/2\pi$ up to $10$ MHz and $\delta/2\pi$ within $(-5,5)$ MHz. To take into account the conditions of experiment Ref. \cite{NV_CPWG_exp} we use $\kappa_b/2\pi=0.4$ MHz, while the effective loss of optical modes are chosen to be the same $\kappa_a=\kappa_b$,
which corresponds to $\gamma_A/2\pi\sim 40$ MHz for the effective decay rate of the excited N-V center, which is reasonable with the experiments \cite{NVlinewidth}.

The quantum master equation we have used can be reduced from a more general Bloch-Redfield master equation \cite{masterEqnBloch,masterEqnRedfield} under low temperature conditions \cite{reducedMasterEqn}. 

\section{Conclusion}\label{sec:concl}
Summarizing, we have investigated quantum statistical properties of optical radiation from a $\Delta$-type cyclic atomic ensembles placed in a microwave cavity and coupled to a two-level atom. In particular, quantum coherence and quantum correlations of the emitted optical photons are examined. Effective models describing the system analogous to coupled driven, dissipative 
linear and nonlinear resonators are obtained under Fr\"ochlich transformations. 

It is shown that analogous to the adiabatic transfer protocol between stationary qubits \cite{darkstate}, the quantum information can be transferred directly between the stationary qubit and the optical flying qubit. The proposed system completes the quantum memory and quantum hardware state transfer protocols by interfacing them with quantum communication channels. In addition, it is found that quantum coherence characteristics can be transferred between the microwave and optical fields.

Furthermore, it is shown that distant entanglement between two optical modes can be realized in the case of two-cyclic atomic ensembles. In addition conditions of three-mode entanglement are also revealed.Controllable particle and mode entanglement for the optical modes or between the optical and the microwave photons is shown to be realizable on demand. 

Finally, diamond crystals with N-V centers in superconducting coplanar waveguide resonators coupled to transmon qubits are proposed to physically realize the model system. The parameters used in numerical simulations are justified for the proposed system. Under strong electric field, it is described that a single N-V center allows for a particular hexagon-type cyclic transitions. Using group contraction method for the ensemble case, the transition type reduces to a diamond-type. For a general treatment of the N-V center taking into account all four crystallographic classes, and using generalized multi-level Morris-Shore transformations a further reduction to $\Delta$-type cyclic scheme is obtained. 

We hope our work can inspire and contribute ongoing efforts for interfacing of optical networks to quantum memories and information processing units.

\acknowledgments

\"O. E. M. gratefully acknowledges A. Atac Imamo\u{g}lu for suggesting this problem and for illuminating discussions. Useful discussions with E. Ilgunsatiroglu are acknowledged. This work is supported by D.P.T. 
(T.R. Prime Ministry State Planning Organization) under the 
Project Number 2009K120200.

\appendix
\section{Coefficients of Effective Hamiltonian}
\label{app:effectiveH_coeffs}
Here we list the coefficients in Eq. \ref{eq:effectiveHfull} explicitly.
\begin{eqnarray}
\varpi_a^\prime&=&
=\varpi_a+g_a^2
\sum_{\lambda}\frac{u_\lambda^2}{\varpi_a-\Omega_\lambda},\\
\omega_b^\prime&=&
=\omega_b+g_b^2
\sum_{\lambda}\frac{v_\lambda^2}{\omega_b-\Omega_\lambda},\\
J&=&-\frac{g_{a}g_{b}}{2}\sum_{\lambda}u_\lambda v_\lambda\left(\frac{1}{\omega_b-\Omega_\lambda}+
\frac{1}{\varpi_a-\Omega_\lambda}\right),\\
\Omega_\lambda^\prime&=&\Omega_\lambda-g_a g_b\sum_\lambda 
 (\frac{u_\lambda  v_\lambda}{\omega_b-\Omega_\lambda} 
+\frac{u_\lambda  v_\lambda}{\varpi_a-\Omega_\lambda}),\\
Q&=&-\frac{1}{2}\sum_{\lambda\neq\mu}
(\frac{g_b^2 v_\lambda v_\mu}{\omega_b-\Omega_\mu}+
\frac{g_a^2 u_\lambda u_\mu}{\varpi_a-\Omega_\mu}),\\
G_{\lambda}&=&-gg_b\frac{v_\lambda}{\omega_b-\Omega_\lambda},\\
E_a&=&\Omega_Ag_b\sum_\lambda\frac{u_\lambda^2}{\varpi_a-\Omega_\lambda},\\
E_b^\prime&=&E_b+\Omega_Ag_b\sum_\lambda
\frac{u_\lambda v_\lambda}{\omega_b-\Omega_\lambda},\\
E_{p\lambda}&=&\Omega_Au_\lambda-E_bg_b
\frac{v_\lambda}{\omega_b-\Omega_\lambda},
\end{eqnarray}
where $\lambda,\mu=\pm$. 

In the $\Omega\gg \Delta$ limit, these coefficients can be approximated to be
\begin{eqnarray}
\varpi_a^\prime&\approx&\varpi_a-\frac{1}{2}
\frac{g_a^2}{\Omega^2}(\Delta_a-\frac{\Delta}{4})\\
\omega_b^\prime&\approx&\omega_b
-\frac{1}{2}\frac{g_b^2}{\Omega^2}(\Delta_b+\frac{3\Delta}{4})\\
J&\approx&\frac{g_a g_b}{\Omega}\left(1+\frac{\Delta^2}{16\Omega^2}\right.\nonumber\\
&&+\left.\frac{(\Delta_b+\Delta/2)^2
+8(\Delta_a-\Delta/2)^2}{2\Omega^2}\right)\\
\Omega_\lambda^\prime&\approx&\Omega_\lambda+\frac{g_b^2}{\Omega^2}
\left(\Delta_b+\frac{3\Delta}{4}\right)
+
\frac{g_a^2}{\Omega^2}
\left(\Delta_a-\frac{\Delta}{4}\right)\\
Q&\approx&-\frac{1}{2\Omega^2}\left [g_a^2\Delta_a-g_b^2\Delta_b-\frac{\Delta}{2}(g_a^2+g_b^2)\right ]\\
G_\pm&\approx&\pm\frac{gg_b}{\sqrt{2}\Omega}\left(1\pm\frac{\Delta_b+3\Delta/4}
{\Omega}\right).\\
E_a&=&-\frac{\Omega_Ag_b}{\Omega^2}(\Delta_a-\frac{\Delta}{4}),\\
E_b^\prime&=&E_b-
\frac{\Omega_bg_b}{\Omega}
\left[1-\frac{\Delta^2+16(\Delta_b+\Delta/2)^2}{16\Omega^2}\right]\\
E_{p\pm}&=&\frac{\Omega_A}{\sqrt{2}}(1\mp\frac{\Delta}{4\Omega})
\mp\frac{E_bg_b}{\sqrt{2}\Omega}
\left(1\pm\frac{\Delta_b+3\Delta/4}{\Omega}\right)
\end{eqnarray}
\section{Dark quasi-spin wave modes}
\label{app:dark_modes}
Here we list the dark quasi-spin wave modes that are uncoupled from the
N-V ensemble Hamiltonian
\begin{eqnarray}
B^\prime&=&\frac{g_{B34}}{g_B g_{B12}}(g_{B1}B_1
+g_{B2}B_2)\nonumber\\
&&-\frac{g_{B12}}{g_B g_{B34}}(g_{B3}B_3
+g_{B4}B_4)\\
B^{\prime\prime}&=& \frac{g_{B2}}{g_{B12}}B_1-\frac{g_{B1}}{g_{B12}}B_2\\
B^{\prime\prime\prime}&=&\frac{g_{B4}}{g_{B34}}B_3-\frac{g_{B3}}{g_{B34}}B_4,
\end{eqnarray}
where we define
\begin{eqnarray}
g_{B12}&=&\sqrt{g_{B1}^2+g_{B2}^2},\\
g_{B34}&=&\sqrt{g_{B3}^2+g_{B4}^2}.
\end{eqnarray}
The same definitions and dark modes for the $A$-mode can be simply obtained by replacing $B\rightarrow A$ in these equations.

\begin{thebibliography}{10}

\bibitem{qMemory}
C.~H. van~der Wal {\em et~al.},
\newblock Science {\bf 301}, 196 (2003).

\bibitem{qProcessor}
M.~Nielsen and I.~L. Chuang,
\newblock {\em Quantum Computation and Quantum Information} (Cambridge Univ.
  Press, Cambridge, UK, 2000).

\bibitem{qComm}
A.~Kuzmich {\em et~al.},
\newblock Nature {\bf 423}, 731 (2003).

\bibitem{hybridDeviceReview}
M.~Wallquist, K.~Hammerer, P.~Rabl, M.~Lukin, and P.~Zoller,
\newblock Phys. Scr. {\bf T137}, 014001 (2009),
\newblock Proceedings of the Nobel symposium on Qu-bits for Quantum
  Information, Gothenburg 2009.

\bibitem{DeltaScheme}
Y.~X.~Liu, J.~Q. You, L.~F. Wei, C.~P. Sun, and F.~Nori,
\newblock Phys. Rev. Lett. {\bf 95}, 087001 (2005).

\bibitem{cyclic_ensemble_model}
Y.~Li, L.~Zheng, Y.-X. Liu, and C.~P. Sun,
\newblock Phys. Rev. A {\bf 73}, 043805 (2006).

\bibitem{hybridSpinEnsembleQubit}
A.~Imamo{\u g}lu,
\newblock Phys. Rev. Lett. {\bf 102}, 083602 (2009).

\bibitem{DeltaArtifAtom}
J.~Siewert, T.~Brandes, and G.~Falci,
\newblock Phys. Rev. B {\bf 79}, 024504 (2009).

\bibitem{DeltaSemiCond1}
G.~Kurizki, M.~Shapiro, and P.~Brumer,  
\newblock Phys. Rev. B {\bf 39}, 3435 (1989).

\bibitem{DeltaSemiCond2}
E.~Dupont, P.~B. Corkum, H.~C. Liu, M.~Buchanan, and Z.~R. Wasilewski,
\newblock Phys. Rev. Lett. {\bf 74}, 3596 (1995).

\bibitem{DeltaSemiCond3}
R.~Atanasov, A.~Hache, J.~L.~P. Hughes, H.~M. van Driel, and J.~E. Sipe,
\newblock Phys. Rev. Lett. {\bf 76}, 1703 (1996).

\bibitem{DeltaChiral1}
P.~Kr{\'a}l, I.~Thanopulos, and M.~Shapiro,
\newblock Phys. Rev. A {\bf 72}, 020303 (2005).

\bibitem{DeltaChiral2}
Y.~Li and C.~Bruder,
\newblock Phys. Rev. A {\bf 77}, 015403 (2008).

\bibitem{magneticdipoleDelta1}
D.~V. Kosachiov, B.~G. Matisov, and Y.~V. Rozhdestvensky,
\newblock I. Phys. B: At. Mol. Opt. Phys. {\bf 25}, 2473 (1992).

\bibitem{magneticdipoleDelta2}
M.~Fleischhauer, R.~Unanyan, B.~W. Shore, and K.~Bergmann,
\newblock Phys. Rev. A {\bf 59}, 3751 (1999).

\bibitem{magneticdipoleDelta3}
B.~Jungnitsch and J.~Evers,
\newblock Phys. Rev. A {\bf 78}, 043817 (2008).

\bibitem{DeltaSchemeQCorrelations}
Y.~H. Ma, Q.~X. Mu, and L.~Zhou,
\newblock International Journal of Theoretical Physics {\bf 46}, 3242 (2007).

\bibitem{chargeQubit}
Y.~Makhlin, G.~Schon, and A.~Shnirman,
\newblock Rev. Mod. Phys. {\bf 73}, 357 (2001).

\bibitem{cpwg_exp1}
A.~Blais, R.-S. Huang, A.~Wallraff, S.~M. Girvin, and R.~J. Schoelkopf,
\newblock Phys. Rev. A {\bf 69}, 062320 (2004).

\bibitem{cpwg_exp2}
A.~Wallraff {\em et~al.},
\newblock Nature (London) {\bf 431}, 162 (2004).

\bibitem{cpwg_exp3}
D.~I. Schuster {\em et~al.},
\newblock Nature (London) {\bf 445}, 515 (2007).

\bibitem{qMemTLR}
P.~Rabl {\em et~al.},
\newblock Phys. Rev. Lett. {\bf 97}, 033003 (2006).

\bibitem{spinsqzParam1}
M.~Kitagawa and M.~Ueda,
\newblock Phys. Rev. A {\bf 47}, 5138 (1993).

\bibitem{spinsqzParam2}
A.~S. rensen, L.~M. Duan, I.~Cirac, and P.~Zoller,
\newblock Nature (London) {\bf 409}, 63 (2001).

\bibitem{Wstate}
W.~D{\"u}r, G.~Vidal, and J.~I. Cirac,
\newblock Phys. Rev. A {\bf 62}, 062314.

\bibitem{NV_CPWG_exp}
Y.~Kubo {\em et~al.},
\newblock Strong coupling of a spin ensemble to a superconducting resonator,
\newblock Unpublished, E-print, arXiv:quant-ph/1006.0251v2, 2010.

\bibitem{rubyDiamondExp}
D.~I. Schuster {\em et~al.},
\newblock High cooperativity coupling of electron-spin ensembles to
  superconducting cavities,
\newblock Unpublished, E-print,arXiv:1006.0242v1, 2010.

\bibitem{darkstate}
H.~R. Zhang, Y.~B. Gao, Z.~R. Gong, and C.~P. Sun,
\newblock Phys. Rev. A {\bf 80}, 062308 (2009).

\bibitem{modematching}
L.~M. Duan, J.~I. Cirac, and P.~Zoller,
\newblock Phys. Rev. A {\bf 66}, 023818 (2002).

\bibitem{phase_transform}
C.~Mewes and M.~Fleischhauer,
\newblock Phys. Rev. A {\bf 72}, 022327 (2005).

\bibitem{collectiveExcOps}
C.~P. Sun, Y.~Li, and X.~F. Liu,
\newblock Phys. Rev. Lett. {\bf 91}, 147903 (2003).

\bibitem{hubbard}
J.~Hubbard,
\newblock Proc. Royal Soc. A {\bf 277}, 237 (1964).

\bibitem{bogoliubov}
N.~N. Bogoluibov,
\newblock J. Phys. (USSR) {\bf 11}, 23 (1947).

\bibitem{frochlich1}
H.~Frochlich,
\newblock Phys. Rev. {\bf 79}, 845 (1950).

\bibitem{frochlich2}
J.~Schrieffer and P.~Wolff,
\newblock Phys. Rev. {\bf 149}, 491 (1966).

\bibitem{frochlich3}
S.~Nakajima,
\newblock Adv. Phys. {\bf 4}, 463 (1953).

\bibitem{JCBH}
A.~D. Greentree, C.~Tahan, J.~H. Cole, and L.~C.~L. Hollenberg,
\newblock Nature Phys. {\bf 2}, 856 (2006).

\bibitem{JCDimer}
C.~D. Ogden, E.~K. Irish, and M.~S. Kim,
\newblock Phys. Rev. A {\bf 78}, 063805 (2008).

\bibitem{qotoolbox}
S.~Tan,
\newblock J. Opt. B: Quantum Semiclass. Opt. {\bf 1}, 424 (1999).

\bibitem{atomPolariton}
S.~Rebi{\'{c}}, S.~M. Tan, A.~S. Parkins, and D.~F. Walls,
\newblock J. Opt. B: Quantum Semiclass. Opt. {\bf 1}, 490 (1999).

\bibitem{GenuineModeEnt}
M.~Hillery and M.~S. Zubairy,
\newblock Phys. Rev. A {\bf 74}, 032333 (2006).

\bibitem{photonBlockade}
M.~J. Werner and A.~Imamo{\u g}lu,
\newblock Phys. Rev. A {\bf 61}, 011801 (1999).

\bibitem{photonblockade2}
A.~Imamo{\u g}lu, H.~Schmidt, G.~Woods, and M.~Deutsch,
\newblock Phys. Rev. Lett. {\bf 79}, 1467 (1997).

\bibitem{spinsqzDissipation}
Y.~Li, Y.~Castin, and A.~Sinatra,
\newblock Phys. Rev. Lett. {\bf 100}, 210401 (2008).

\bibitem{qOpticsBook}
M.~Scully and M.~S. Zubairy,
\newblock {\em Quantum Optics} (Cambridge University Press, Cambridge, 1997).

\bibitem{NtypeAtom}
S.~Rebic, J.~Twamley, and G.~J. Milburn
\newblock Phys. Rev. Lett. {\bf 103}, 150503 (2009).

\bibitem{CPT}
H.~R. Gray, R.~Whitley, and J.~C.~R. Stroud,
\newblock Opt. Lett. {\bf 3}, 218 (1978).

\bibitem{GHZstate1}
D.~M. Greenberger, M.~Horne, and A.~Zeilinger,
\newblock in {\em Bell's Theorem, Quantum Theory, and Conceptions of the
  Universe}, edited by M.~Kafatos, p.~69, Dordrecht, 1989, Kluwer.

\bibitem{GHZstate2}
D.~Bouwmeester, J.-W. Pan, M.~Daniell, H.~Weinfurter, and A.~Zeilinger,
\newblock Phys. Rev. Lett. {\bf 82}, 1345 (1999).

\bibitem{distillEPRfromW}
B.~Fortescue and H.-K. Lo,
\newblock Phys. Rev. Lett. {\bf 98}, 260501 (2007).

\bibitem{transmon}
J.~M. {\it et al.},
\newblock Nature (London) {\bf 449}, 443 (2007).

\bibitem{transmonTheory}
J.~Koch {\em et~al.},
\newblock Phys. Rev. A {\bf 76}, 042319 (2007).

\bibitem{transmonExp}
J.~A. Schreier {\em et~al.},
\newblock Phys. Rev. B {\bf 77}, 180502 (2008).

\bibitem{CooperPairBoxJCM}
A.~Blais, R.~S. Huang, A.~Wallraff, S.~M. Girvin, and R.~J. Schoelkopf,
\newblock Phys. Rev. A {\bf 69}, 062320 (2004).

\bibitem{NVLevelsTheory}
A.~Lenef and S.~C. Rand,
\newblock Phys. Rev. B {\bf 53}, 13441 (1996).

\bibitem{NVLevelsExp}
A.~Lenef, S.~W. Brown, D.~A. Redman, S.~C. Rand, J.~Shigley, and E.~Fritsch,
\newblock Phys. Rev. B {\bf 53}, 13427 (1996).

\bibitem{spinflipExp1}
C.~Santori {\em et~al.},
\newblock Optics Express {\bf 14}, 7986 (2006).

\bibitem{SpinFlipExp2}
P.~R. Hemmer, A.~V. Turukhin, M.~S. Shahriar, and J.~A. Musser,
\newblock Optics Letters {\bf 26}, 361 (2001).

\bibitem{spinflipNVCexp}
P.~Tamarat {\em et~al.},
\newblock New J. Phys. {\bf 10}, 045004 (2008).

\bibitem{spinPreservingExp}
F.~Jelezko {\em et~al.},
\newblock Appl. Phys. Lett. {\bf 81}, 2160 (2002).

\bibitem{lambdaNVbyE}
G.~Gonz\'{a}lez and M.~N. Leuenberger,
\newblock Phys. Rev. B {\bf 80}, 201201(R) (2009).

\bibitem{morrisShore1}
J.~R. Morris and B.~W. Shore,
\newblock Phys. Rev. A {\bf 27}, 906 (1983).

\bibitem{morrisShore2}
A.~A. Rangelov, N.~V. Vitanov, and B.~W. Shore,
\newblock Phys. Rev. A {\bf 74}, 053402 (2006).

\bibitem{morrisShore3}
F.~E. Zimmer, J.~Otterbach, R.~G. Unanyan, B.~W. Shore, and M.~Fleischhauer,
\newblock Phys. Rev. A {\bf 77}, 063823 (2008).

\bibitem{striplineCavityLoss}
A.~Blais {\em et~al.},
\newblock Phys. Rev. A {\bf 75}, 032329 (2007).

\bibitem{NVlinewidth}
P.~Tamarat {\em et~al.},
\newblock Phys. Rev. Lett. {\bf 97}, 083002 (2006).

\bibitem{masterEqnBloch}
F.~Bloch,
\newblock Phys. Rev. {\bf 05}, 1206 (1957).

\bibitem{masterEqnRedfield}
A.~G. Redfield,
\newblock IBM J. Res. Develop. {\bf 1}, 19 (1957).

\bibitem{reducedMasterEqn}
I.~Rau, G.~Johansson, and A.~Shnirman,
\newblock Phys. Rev. B {\bf 70}, 054521 (2004).

\end{thebibliography}



\end{document}